\documentclass{elsart}
\usepackage{epsfig}
\usepackage{cite}

\def\Journal#1#2#3#4{{#1} {\bf #2}, #3 (#4)}
\def\NIM{\em Nucl. Instrum. Methods}
\def\NIMA{{\em Nucl. Instrum. Methods} A}

\newcommand{\lsim}{\buildrel < \over {_\sim}}
\newcommand{\gsim}{\buildrel > \over {_\sim}} 
\begin{document}
\begin{frontmatter}
\title{Beam Test of Silicon Strip Sensors 
       for the ZEUS Micro Vertex Detector}
\vspace*{0.5cm} 
\mbox{L.A.T.~Bauerdick~$^{1 (a)}$, E.~Borsato~$^{2}$,  C.~Burgard~$^{1}$, T.~Carli~$^{1}$, R.~Carlin~$^{2}$,} 
\mbox{M.~Casaro~$^{2}$, V.~Chiochia~$^{1}$, F.~Dal Corso~$^{2}$, D.~Dannheim~$^{1}$, A.~Garfagnini~$^{5 (b)}$,} 
\mbox{A.~Kappes~$^{3}$, R.~Klanner~$^{5}$,  E.~Koffeman~$^{4}$, B.~Koppitz~$^{5}$, U.~K\"otz~$^{1}$, E.~Maddox~$^{4}$,} 
\mbox{M.~Milite~$^{1 (c)*}$, M.~Moritz~$^{1}$, J.S.T.~Ng~$^{1 (d)}$, M.C.~Petrucci~$^{1}$, I.~Redondo~$^{6 (e)*}$,}  
\mbox{J.~Rautenberg~$^{3 (f)}$, H.~Tiecke~$^{4}$,  M.~Turcato~$^{2}$, J.J.~Velthuis~$^{4}$, A.~Weber~$^{3}$}
\\
\vspace*{0.4cm}
\small{
\begin{enumerate}
\item  Deutsches Elektronen-Synchrotron DESY, Hamburg, Germany
\vspace*{-0.3cm}\item  Dipartimento di Fisica dell' Universit\`a and INFN, Padova, Italy
\vspace*{-0.3cm}\item  Physikalisches Institut der Universit\"at Bonn, Bonn, Germany
\vspace*{-0.3cm}\item  NIKHEF and University of Amsterdam, Amsterdam, Netherlands
\vspace*{-0.3cm}\item  Hamburg University, Institute of Exp. Physics, Hamburg, Germany
\vspace*{-0.3cm}\item Universidad Autonoma de Madrid, Madrid, Spain 
\vspace*{-0.3cm}\item[(a)] Now at Fermi National Accelerator Laboratory FNAL, Batavia, Illinois, USA  
\vspace*{-0.3cm}\item[(b)] Now at Dipartimento di Fisica dell' Universit\`a and INFN, Padova, Italy
\vspace*{-0.3cm}\item [(c)]Now at Hamburg University, Institute of Exp. Physics, Hamburg, Germany
\vspace*{-0.3cm}\item [(d)]Now at the Stanford Linear Accelerator Center  SLAC, Stanford, California, USA  
\vspace*{-0.3cm}\item [(e)] Now at Laboratoire Leprince Ringuet - Ecole Polytechnique, Route de Saclay 91128 Palaiseau Cedex, France  
\vspace*{-0.3cm}\item [(f)] Supported by the GIF, contract I-523-13.7/97.
\vspace*{-0.3cm}\item [*] {Corresponding authors. Tel: +33 1 69 33 44 05
 ; fax:+33 1 69 33 30 02 \\
 $E{\rm-}mail~addresses:$ redondo@poly.in2p3.fr (I.~Redondo), milite@mail.desy.de (M.~Milite).}
\end{enumerate}

}

\begin{abstract}
For the HERA upgrade, the ZEUS experiment has designed and installed a high precision Micro Vertex Detector
(MVD) using single sided $\mu$-strip sensors with capacitive charge division. The sensors have a readout 
pitch of 120 $\mu$m, with five intermediate strips (20 $\mu$m strip pitch). An extensive test program has 
been carried out at the DESY-II testbeam facility.  In this paper we describe the setup developed to test the 
ZEUS MVD sensors and the results obtained  on both irradiated and non-irradiated single  sided
$\mu$-strip detectors with rectangular and trapezoidal  geometries.
The  performances of the sensors coupled to the readout electronics (HELIX chip, version 2.2)  have been 
studied in detail,  achieving a good description by a Monte Carlo simulation.
Measurements of the position resolution as a function of the angle of incidence are presented, focusing 
in particular on the comparison between standard and newly developed reconstruction algorithms.

$PACS:$  29.40.Gx; 29.40.Wk; 07.05.kj

$Keywords:$ ZEUS; Beam test; Silicon; Microstrip; Position reconstruction algorithms

\end{abstract}
\end{frontmatter}


\section{Introduction}

The HERA $e p$ collider luminosity upgrade~\cite{bib:hera_upgrade_scheek}  performed 
during the years 2000-2001 aims to  increase the instantaneous luminosity
from 1.5 to $6\cdot 10^{31}$~cm$^{-2}$ s$^{-1}$, providing thus a
higher sensitivity to low cross section physics.
The ZEUS experiment~\cite{bib:zeus_det} has been equipped with a new silicon
Micro Vertex Detector (MVD) which is going to improve the global
precision of the existing tracking system, allowing to
identify events with secondary vertices coming from the decay of
long-lived states such as  hadrons with charm or bottom  and $\tau$ leptons.
Moreover, the detector acceptance will be enhanced in the forward region, along
the proton beam direction, improving for 
 example
 the detection of very
high $Q^2$
scattered electrons 
and the reconstruction of the interaction vertex 
in  high $x$ 
 charged current events\footnote{ $-Q^2$ is the exchanged photon invariant mass; $x$  is the fraction of 
the proton momentum carried by the struck quark.}.

\begin{figure}
\begin{center}
\epsfig{file=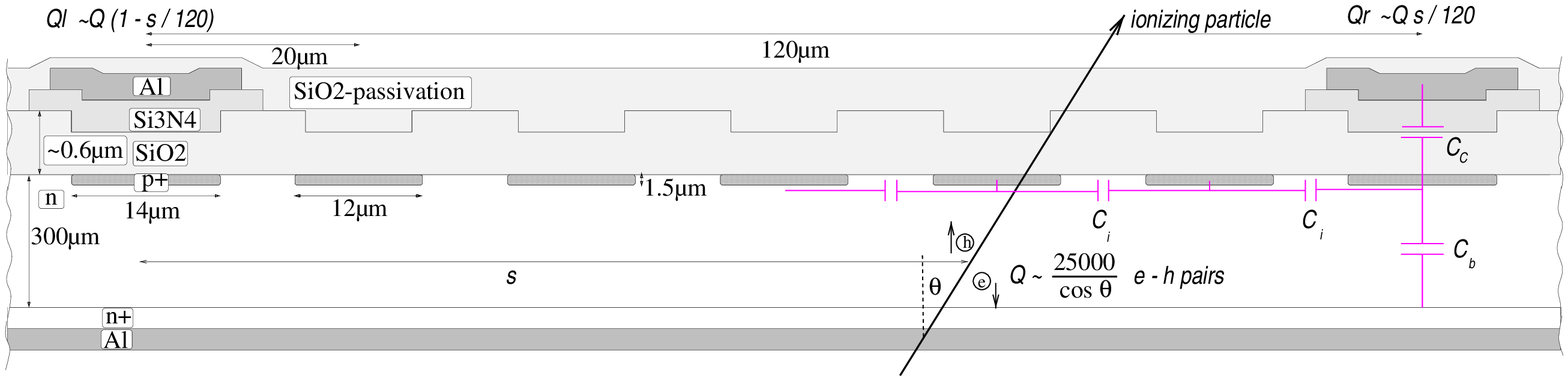,width=1.\linewidth,height=5cm}
\end{center}
\caption{\label{fig:mvddet} Cross section of a MVD silicon sensor between two readout strips (drawing not to scale).
 Dimensions of the layers are given in the left part of the picture. On the right side a  simplified picture of the capacitive network is shown.} 
\end{figure}

%
According to the design specifications~~\cite{bib:mvd_mech_paper10,bib:mvd_mech_paper9,bib:mvd_mech_paper6,bib:mvd_mech_paper5}, the MVD
is composed of a barrel (BMVD) and forward (FMVD) part, requiring a
good matching with the existing detectors.
The MVD had to fit inside a cylinder of 324 mm diameter defined by the inner wall 
of the Central Tracking Detector (CTD). The readout electronics, based on the
HELIX  chip (version 3.0)~\cite{bib:helix_ref1,bib:helix_ref2} is mounted inside the active
area, close to the silicon diodes.
The silicon sensors are single sided, AC coupled, strip detectors with
capacitive charge division; the readout pitch is $120~\mu$m and the strip
pitch is $20~\mu$m.
A sketch of the silicon sensor can be seen in figure~\ref{fig:mvddet}. 
 The BMVD  (FMVD-1,FMVD-2) sensors have a rectangular (trapezoidal) geometry 
 with  6.4 cm x 6.4 cm  
(base=6.4 cm $\times$ height=7.35 cm and 6.4 cm $\times$4.85 cm, respectively)
 dimensions. 
 In the FMVD,  the sides  of the trapezoid are tilted by 180$^{\circ}$/14 with respect to the 
bases and  the strips are parallel  to one of the tilted sides of the trapezoid, having thus 
different lengths  across the sensor.  
The biasing of the strips is implemented using poly-silicon resistors ($\sim$ 1.5-2.5 M$\Omega$) 
connected to  the sensor  ground line (called in the following biasing ring),
located alternatively on both ends of the strips. 
The first and last strips close the biasing ring, being  directly connected 
to it. Three p$^{+}$ guard rings, designed to adjust the potential towards the 
detector edges, surround the sensitive area. An additional n$^{+}$ doped implant beyond the last guard ring
allows to bias the backplane with a contact from the top. 
Detailed descriptions of the MVD design and mechanical structure can be
found in~\cite{bib:mvd_mech_paper6,bib:mvd_mech_paper3,bib:mvd_mech_paper2,bib:mvd_mech_paper1}. 
The detailed design of the silicon sensors and results on the electrical measurements are described 
elsewhere~\cite{bib:mvd_mech_paper7,bib:mvd_mech_paper4,bib:mvd_elec_paper}.

The testbeam program for the MVD had several goals:
\begin{itemize}
  \item to study the general performance of sensors with minimum
           ionising particles (i.e. noise level, pedestal stability, hit efficiency, charge division);
  \item to test prototype versions of frontend electronics (i.e. readout chips and hybrids) ;
  \item to test the sensors at different bias voltages;
  \item to measure the position resolution for different angles of incidence
        in order to optimise reconstruction algorithms;
   \item to study the effect of irradiation. 
\end{itemize}
The data have been compared with the results of a  Monte Carlo program for the silicon sensor 
simulation in order to gain input for the vertex detector simulation.
The detector performances have been studied using three non-irradiated and two additional 
irradiated  sensors~\cite{bib:mvd_mech_paper8,bib:mvd_elec_paper}. 
A barrel sensor was irradiated using photons from a  $^{60}$Co source:  
the detector was left floating and irradiated up to an integrated dose of 2.0 kGy. 
A second sensor was irradiated with reactor neutrons having a fluency $ \phi_{e} = 10^{13}$   
1~MeV$_{\rm equiv.}$ n/cm$^{2}$. No substantial effects due to radiation damage on the detector 
performances  have been observed.
All  results presented in the following sections  refer to non-irradiated sensors  and to the sensor 
irradiated floating with 2~kGy of $^{60}$Co photons. 

After a  brief description of the testbeam setup in section ~\ref{sec:setup},
the treatment of the data and the general  performance of the detectors are summarised 
in section~\ref{sec:data_analysis}. The detector simulation is described in section~\ref{sec:simulation}.
Section ~\ref{sec:ene_reso} is devoted to studies which   use perpendicular tracks. It also describes 
the  extraction of the intrinsic position resolution. The position resolution as a function of the angle of incidence
is  studied in detail in sections~\ref{sec:smallangle} and ~\ref{sec:ang_reso}.
The paper ends with a summary of the results.

\section{Test beam setup}
\label{sec:setup}
The measurements were performed at the DESY-II testbeam,
a parasitic electron beam obtained after two conversions: a $10~\mu$m thick 
carbon-fiber target in the machine intercepts the beam and produces 
bremsstrahlung photons which are  converted into electron-positron pairs 
in a 0.1 X$_0$ thick copper target.
 A bending magnet together with a momentum
defining collimator slit delivers the beam into the experimental
hall. Depending on the primary use of DESY-II the maximal momentum
varies between 4.3 and 7.5 GeV/c. Most measurements were done at 3 and
6 GeV/c resulting in a trigger rate of $\sim$ 10 Hz and $\sim$ 2 Hz, respectively. 

\begin{figure}
\begin{center}
\epsfig{file=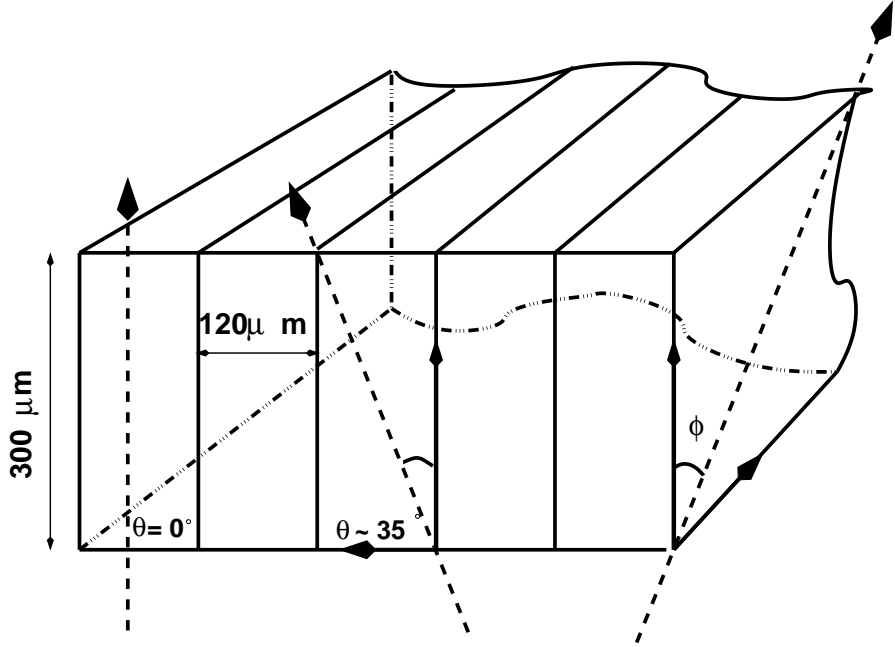,width=12cm}
\end{center}
\caption[Schematic cross section of a MVD silicon sensor to scale]{\label{fig:detec_onscale} Schematic cross 
section of a MVD silicon sensor to scale. The angles of incidence $\theta$ and $\phi$ are indicated.} 
\end{figure}

\begin{figure}
\begin{center}
\epsfig{file=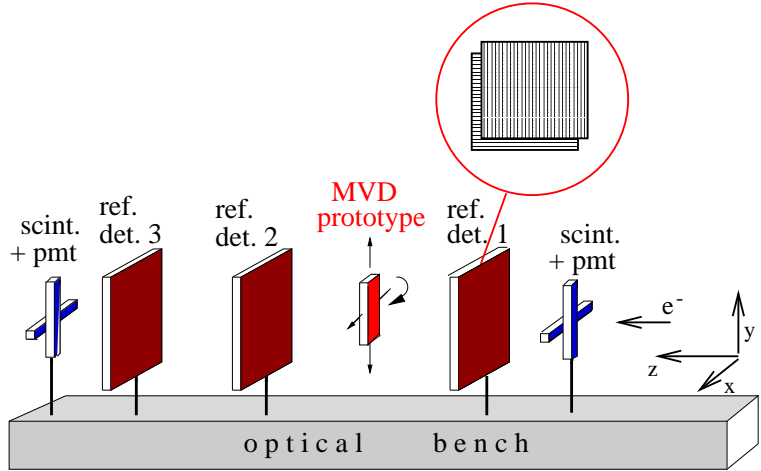,width=10cm}
\end{center}
\caption[Schematic drawing of the testbeam setup]{\label{fig:tbscetch} Schematic drawing of the testbeam setup. } 
\end{figure}

A silicon reference telescope has been assembled to allow a precise determination of the particle impact point 
on the detector to be studied. Both, the telescope modules and the module holding  the MVD detector 
 are mounted on a common optical bench. The detector to be studied is mounted
between the telescope modules on linear and rotational positioners which
allow to investigate the performance in different areas of the detector and for
different angles of incidence. The rotations can be
 around the strip axis ($\theta$ angle) or around the axis perpendicular to the strip in  
the detector plane ($\phi$ angle) (see figure~\ref{fig:detec_onscale}).
A trigger was generated by  coincidence of
the signals from 
scintillator fingers located at both end sides of the optical bench.
A drawing of the testbeam  setup is shown in figure~\ref{fig:tbscetch}.
The data acquisition system is based on an  embedded Sun
workstation in a VME-crate which controls the initialisation of the
readout modules, the readout chips, the pattern generators, the
GPIB-interface for the positioners and the data taking. The system
is run under LabView~\cite{bib:labview}.

\subsection{Beam telescope}
\label{ss:telescope}
The beam telescope consists of 3 modules which define the track
in two coordinates along the telescope. The modules are a  
version of a CERN development described in~\cite{bib:telescope,bib:tele_res}.
They consist
of two $300~\mu$m thick single-sided silicon detectors of $32 \times 
32$~mm$^2$ size
with a  strip
 pitch of $25~\mu$m and a readout pitch of $50~\mu$m;
 the strip direction of the detectors  are 
perpendicular to each other.

All detectors have a very good S/N (i.e. signal over noise ratio for minimum ionising particles)\footnote{The term 
noise (N) refers to the single strip noise.},  $80 < S/N < 130$, and a low fraction of dead channels and 
noisy strips ($< 0.1\%$).
The extrapolation to infinite momentum results in an intrinsic resolution 
of $2.8 \pm 0.1~\mu$m (in  section~\ref{sec:ene_reso} the analysis method 
applied to determine the intrinsic resolution of a silicon detector is discussed).

 The MVD detector is  aligned with respect to the closest reference detectors (1 and 2). 
 The  achieved precision of the alignment procedure~\cite{bib:margherita_phd} is $<$\,1~$\mu$m.

\subsection{MVD detector readout}
The MVD detectors were operated in the test beam using prototype versions
of the HELIX readout chip (version 2.2), developed 
for the Hera-B experiment~\cite{bib:helix_ref1,bib:helix_ref2}.
The HELIX chip provides 128 analog
channels with a charge sensitive pre-amplifier and shaper, forming a semi-Gaussian pulse
 with a peaking time of $\sim$ 50-70 ns. The signals are sampled in an analog pipeline capable 
of storing 128 events with 8 extra channels for trigger derandomisation. 
The chip can be operated up to 40 MHz clock-rate. In the ZEUS
experiment it will be used at 10 MHz write and read speed. In order
to simplify the readout system in the testbeam, it was decided to
synchronise the chip clock to the synchrotron revolution frequency of
about 1.05 MHz; with this setting particles traversing the detectors were in a
fixed phase relation to the chip clock. 
Tests performed using a frequency multiplier of 10 (which are not discussed in the present paper) 
showed no difference in the detector performances.
     
\subsection{MVD detector  assembly}

In the testbeam, a protection circuit,  including  a
protection resistance of 0.5 M$\Omega$, was introduced  between the backplane contact 
and the power supply. In the following, the  voltage applied between the biasing ring and the 
backplane  is referred to as V$_{bias}$. 
The detectors were biased at full depletion, unless otherwise stated. 
The Barrel and Forward MVD detectors are connected to the front-end electronics
using a Upilex~\cite{bib:upilex}  ``fan-out'' foil with conductive lines. 
The Upilex circuits are made of an Upilex S substrate, a 50 $\mu$m thick
polyimide film. A conductive layer of 5 $\mu$m electro-plated copper is
deposited on top and separated from the substrate by means of a 150 nm
thick nickel adhesion layer. A 1.5 $\mu$m gold layer is deposited over the
conductive strips and the pads used for bonding.
The Upilex circuits for the Barrel and Forward modules are produced at
CERN~\cite{bib:Gandi}.
The strip pitch is $120~\mu$m on the detector side and is reduced to
$100~\mu$m on the hybrid, where the Upilex strips are connected to a pitch
adapter which further reduces the readout pitch to $41.4~\mu$m of the HELIX
input bond pads.
The front-end electronic is mounted on a multi-layer Hybrid structure
($40 \times 70~\mathrm{cm}^2$) supporting 4 HELIX chips which are needed
to read-out the 512 strips of a BMVD detector; for the FMVD detector only
480 readout channels are required.

\section{Data analysis and general performance}
\label{sec:data_analysis}
The digitised ADC output coming from the MVD detector and the telescope detectors
is stored in files during the data taking.
The channel noise and pedestal levels are measured using special random
trigger runs of 100-200 events taken without beam.
In order to reduce the data volume, for the telescope data,
zero suppression is performed directly in the CAEN V550 ADC, using 
a threshold level of 3 times the channel noise.
No selection is applied to the raw data for the MVD detector.

During the offline analysis, the common mode noise (CMN) and the pedestal
levels are subtracted from the data~\cite{bib:margherita_phd}.
The pedestal, determined once per day, has shown negligible variation over time.
The variation of the pedestals within one readout chip, from the first to the last readout channel, has been found 
to be of the order of the cluster pulse height. The strip noise was stable and showed uniform behaviour;   
its variation within regions read by the same chip are much smaller than those observed  between chips ($\sim$ 20\%).
No dependence of the noise on the strip length (varying between 6 mm and 73.3 mm) 
was observed~\cite{bib:redondo_phd}. 
The common mode noise is Gaussian distributed with a rms comparable to the single strip noise level.

\begin{figure}
\begin{center}
\epsfig{file=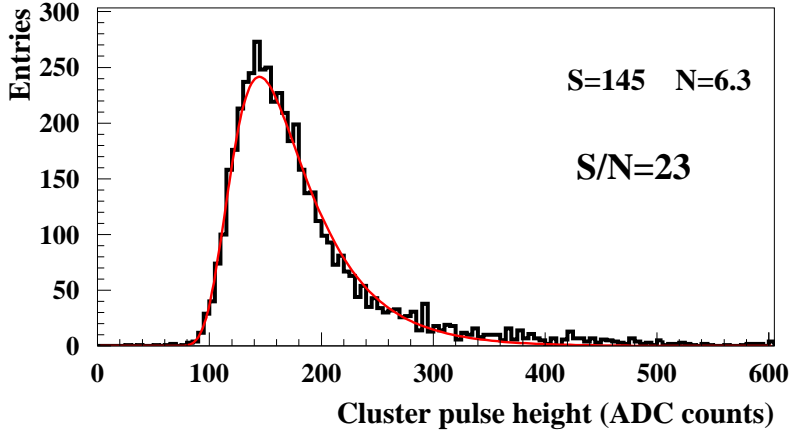,width=12cm}
\end{center}
\caption{Two strip cluster pulse height distribution.  A Landau fit is superimposed to the data; 
the following parameterisation has been used:  
$p_1 \exp{ \left[ -0.5\cdot \left( \lambda + \exp{(-\lambda)} \right) \right]}$  
where $\lambda = p_3 (x-p_2)$, and $p_1$, $p_2$ and $p_3$ are free parameters of the fit.}
\label{fig:sn}
\end{figure}
A cluster seed is identified by looking for the highest signal strip in the detector. All neighbouring strips
with a signal larger than a certain threshold level T (usually T = 3$\times \sigma_{\rm chip}$, where 
$\sigma_{\rm chip}$ is the average chip noise) are added to form a cluster. 
The cluster pulse height  and size are then defined as the sum of the 
signals from all the strips and the total number of strips belonging to the cluster, respectively. 

For the determination of the S/N using perpendicular tracks, only the strip with the highest signal 
and its neighbouring strip (left or right) with the higher pulse height  are selected and the sum of their pulses  defines in 
this case the total cluster signal. Figure~\ref{fig:sn} shows the resulting cluster pulse height distribution: 
the data are fitted by a Landau function. 
A S/N between 20 and 24 has been obtained for different detectors and readout chips.

An asymmetric cross talk has been observed in  the HELIX readout chip:
measurements with an external test pulse have shown that when pulsing
a channel, a fraction of the input charge is found
on the previous (next) channel for even (odd) channels~\cite{bib:padova_helix}.
Using testbeam data, the asymmetric cross talk has been determined to be 
around 5\% for all chips~\cite{bib:margherita_phd}.
The cause of this  effect  is presumably due to an asymmetry in the chip pipeline design.
All testbeam data are corrected for  asymmetric cross talk. 

\subsection{Gain calibration}
\begin{figure}
\begin{center}
\epsfig{file=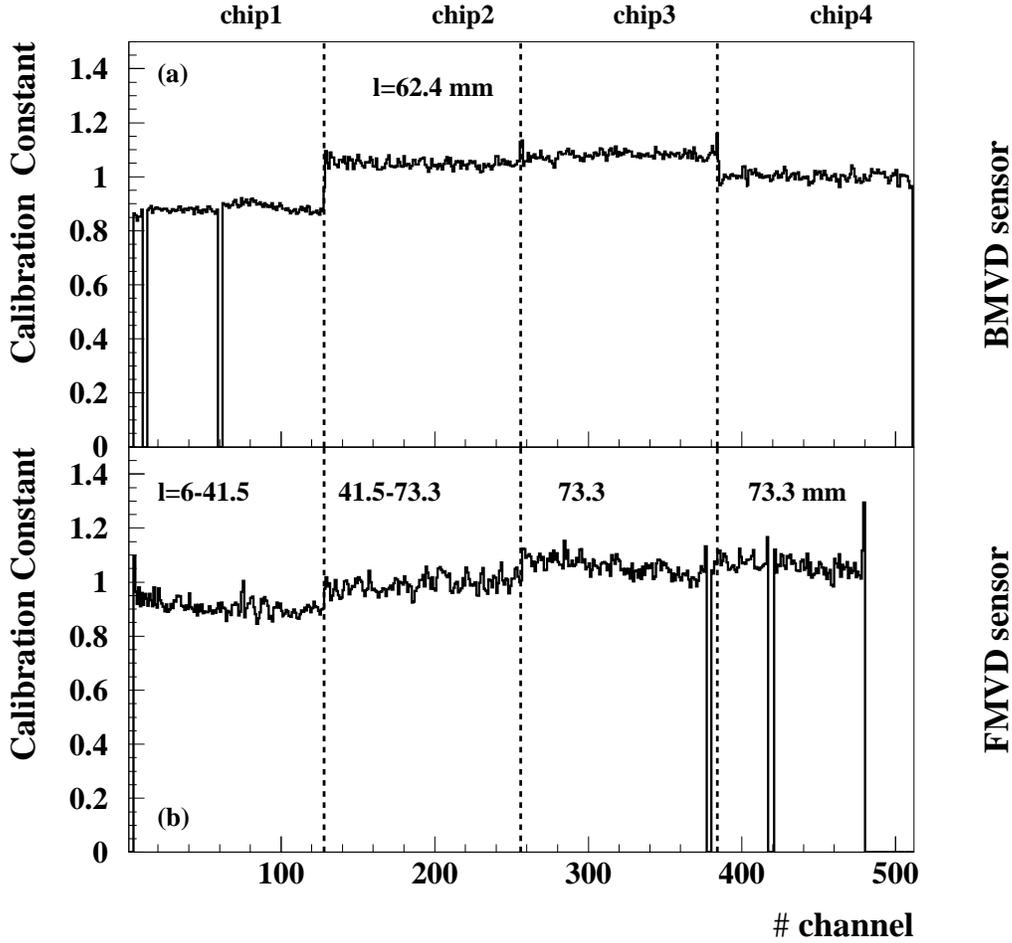 ,width=1\linewidth}
\end{center}
\caption[Strip calibration constants for 512 strips of one detector]{\label{stripcal} 
 Calibration constants for 512 channels of one BMVD detector (a) and 480 of a FMVD-1 detector (b) 
readout by four HELIX chips each. The strip regions read out by different chips are separated 
by the dashed vertical lines. ``l'' is the strip length range for one chip region.}  
\end{figure}

The whole detector was illuminated in order to study the uniformity in gain  of  the  
channels in terms of  the relative calibration constants, cal(i):
\begin{equation}
\hspace{-0.5cm} {\rm cal(i) = \frac{<\Sigma>_{channels}}{\Sigma(i)} \;\;\; {\rm with} \;\;\; 
 \Sigma(i)=\overline{S_{max-1}}^{hit}(i)+\overline{S_{max}}^{hit}(i)+\overline{S_{max+1}}^{hit}(i)}
\end{equation}
where i is the channel number; $< >_{\rm channels}$ is an average over channels
; $\overline{S }^{\rm hit}$ averages over hits;
  $S_{\rm max}$ is the charge of the strip with the  maximum charge of the event 
  and $S_{\rm max\pm1}$ is the charge of the strip with  position 
   $\#{\rm maximum}\pm1$.  The  gain for all channels of a BMVD and a FMVD-1 detector are shown 
in figure~\ref{stripcal}. The mean gain value and the rms of each chip are given in
table \ref{t:scalnum}. The first and the last strip (which have only one
neighbour) as well as all the broken strips and their direct neighbour
strips are excluded  and their calibration constants are set to zero in the calibration procedure. 
 The differences of the strip gains for one chip are smaller than
2\% and comparable with the statistical uncertainty on the gain calibration ($\sim$ 1\%); 
the calibration constants differ from chip to chip by up to 20 \%.
\begin{table} [h]
  \begin{center}
    \caption[Average chip calibration constants and their rms for a BMVD  detector]{\label{t:scalnum}{
S/N, mean and  rms of the  calibration constants in   regions read out by the same chip  for a BMVD and FMVD-1 detectors.}}
    \small
    \begin{tabular}{ l l l l l l l l l} \hline
        & \multicolumn{4}{l}{ BMVD detector} & \multicolumn{4}{l}{ FMVD-1 detector} \\ \cline{2-9}
        & chip 1 & chip 2 & chip 3 & chip 4 &   chip 1 & chip 2 & chip 3 & chip 4 \\ \hline
      length (mm)&62.4 &62.4 & 62.4 & 62.4 & 6-41.5   & 41.5-73.3  &  73.3  & 73.3  \\ \hline
      {Mean} & 0.884 & 1.050 & 1.081  & 1.001   & 0.909   &   0.990      &  1.062  &  1.070 \\ \hline
      {Rms}  & 0.016 & 0.017 & 0.016  & 0.017   & 0.036   &   0.024      &  0.029  &  0.039 \\ \hline
            S/N     &  20.8     &   21.9    &  21.4       &   21.2     & 24.7    &    23.6      &  21.0   &  20.7 \\ \hline
    \end{tabular}
  \end{center}
\end{table} 

 A possible dependence of the gain on the strip length has been investigated since it could 
introduce left-right asymmetries affecting the position reconstruction algorithms.
The rms of the calibration constants  for strips  within a chip of a FMVD-1 detector 
(table~\ref{t:scalnum}, right) are comparable to the statistical uncertainty (2-3 \%).
The mean of the calibration constants and the S/N within a chip follow
the same pattern:
 regions 3  and 4 (long and constant length strips) have smaller  signal than
region 1 and 2 (short and variable length strips). 
No correlation between the calibration  constants and the strip number 
(and therefore the strip length) is observed.
Since the relative gain is constant for a readout chip, no calibration has been applied in the analyses presented in 
this paper\footnote{The scintillator fingers used for triggering define a surface of 9 mm $\times$9 mm which is well within 
the detector area readout by a single HELIX chip.}.

\subsection{Hit efficiency}
The hit efficiency is defined as
\begin{equation}
\rm \epsilon =  N_{events}({\rm Hit})/N^{triggered}_{events}
\end{equation}
where  ``Hit'' is equivalent to the  presence of 1 strip with signal
larger than $n$ times the noise ($n$=5,3) in a region of 2.4 mm centred on 
 the position predicted by the telescope. The particle impact position was reconstructed using two reference 
detectors; in order to avoid contamination from double tracks, only events with a single cluster in all planes of the 
two telescope modules  were  accepted.

Using a sample of $\sim10^5$ events the 90\% confidence limits on $\epsilon$ is  
$99.96\%>\epsilon_5>99.95\% \,(99.997\%)$   for a signal in the MVD detector  larger than 5 (3) times the noise.

\subsection{Detector performance as a function of the bias voltage}
\begin{figure}[hbtp]
\begin{center}
 \mbox{
\epsfig{file=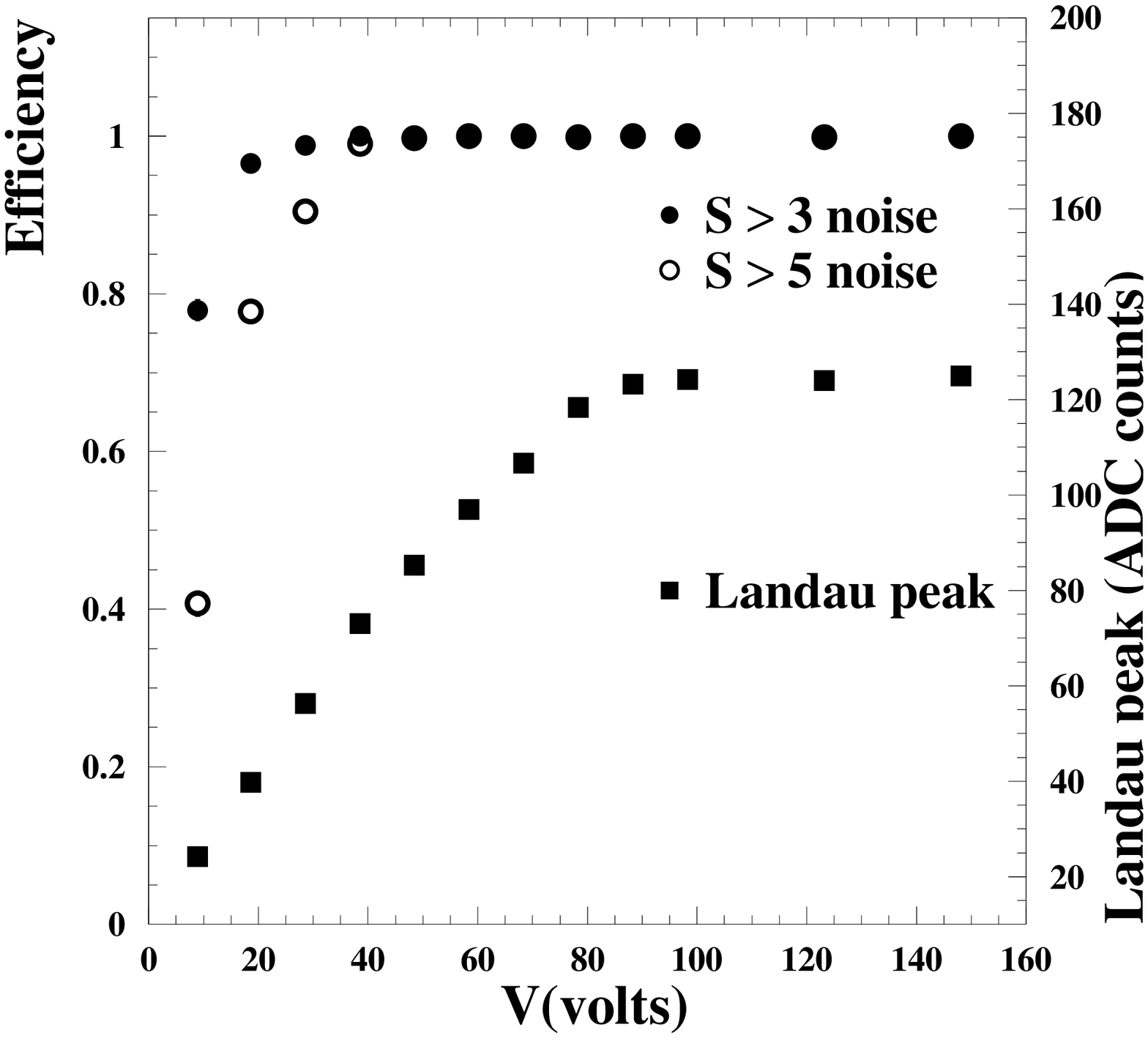,width=0.49\linewidth}
 \put(-1.9,1.8){(a)}
\hspace*{0.6cm}
\epsfig{file=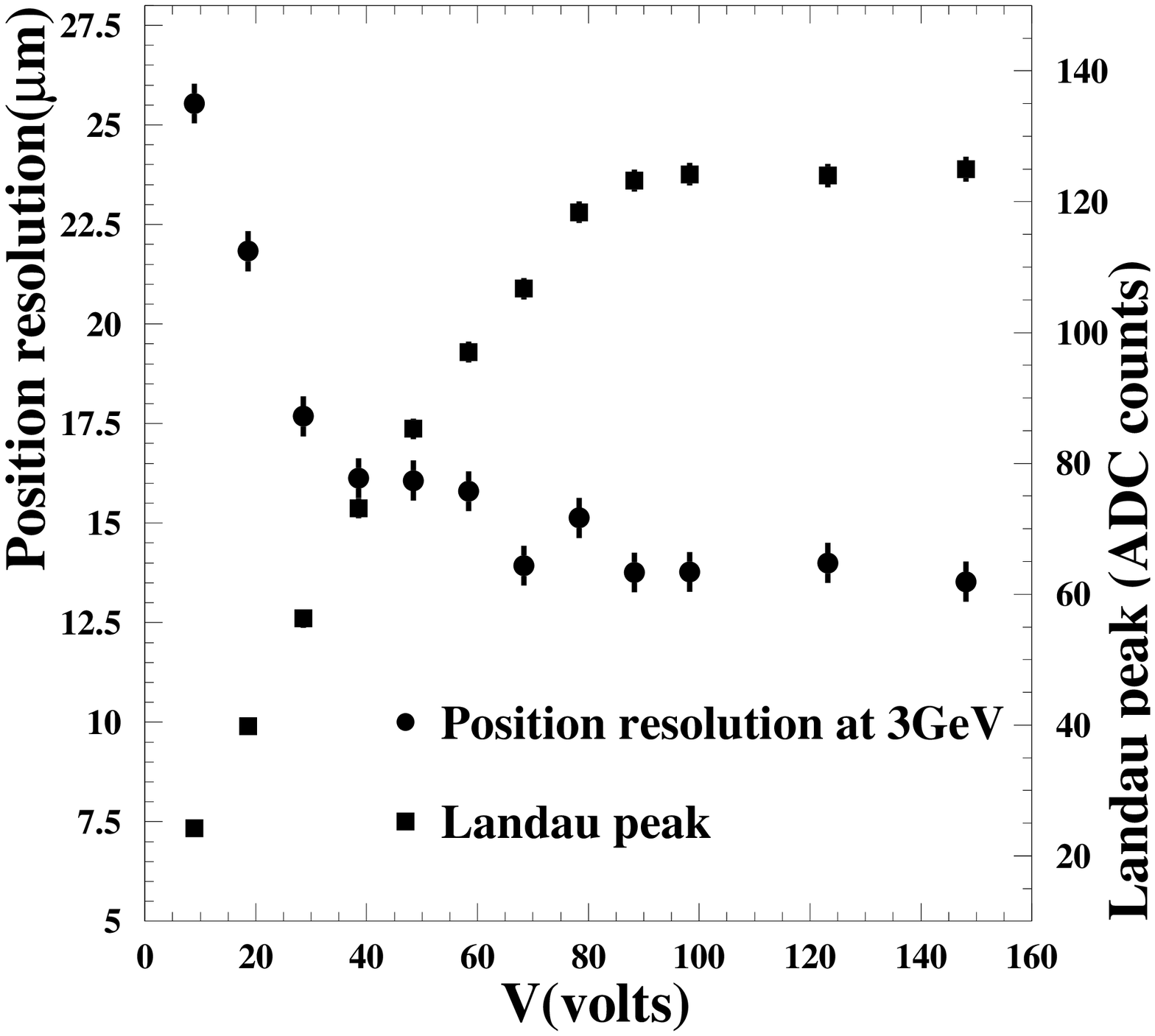,width=0.49\linewidth}
\put(-1.9,1.8){(b)}
}
\end{center}
\caption[$\epsilon$ vs $V_{bias}$]{ (a)  Most probable value of the energy loss distribution (squares) and hit 
efficiency  (dots) as a function of the bias voltage; (b)  position resolution (dots) as a function of the bias voltage.}
\label{fig:eff}
\end{figure}

The dependence of the detector performance  as a function of $V_{bias}$
has been investigated.
Figure ~\ref{fig:eff}(a) shows the most probable value of the energy loss distribution\footnote{The most probable
  value of the energy loss distribution is defined as the peak value from a fit by a Landau function.} 
and the hit efficiency as a function of the bias voltage. For values of $V_{bias}$  above the depletion voltage ($V_{dep} \simeq 85 V$) 
a plateau is reached.  The signal collection degrades with decreasing $V_{bias}$, whereas the hit efficiency 
remains constant at $V_{bias}$ well below the depletion voltage.  Only for $V_{bias}\lsim 40$~V $\epsilon$ 
starts to decline. The noise level remains constant ($\sim 6$ ADC counts) for  values of $V_{bias} > 5$V.

The charge division of the detector and the position resolution (see sections~\ref{sec:charge_sharing} 
and ~\ref{sec:ene_reso}, respectively, for detailed explanations)  were also studied as a function
of $V_{bias}$. Figure~\ref{fig:eff}(b)  shows the position resolution as  a function of $V_{bias}$: a stable 
behaviour above $\sim$ 40 V can be observed. The charge division of the detector has also
found to be unvaried above $\sim$ 40 V~\cite{bib:redondo_phd}. It is remarkable that with $\sim9~V$ of effective
voltage in the detector, with only $\sim 30\%$ of the bulk  depleted,  a resolution of  25 $\mu m$ is achieved.
As a conclusion, the performance of the detector seems to be rather stable well below depletion voltage.

\subsection{Charge division}
\label{sec:charge_sharing}
The charge division of the MVD 
 sensors
 with five
intermediate strips  was studied using
 tracks  perpendicular to the detector plane.

\begin{figure}
\begin{center}
\epsfig{file=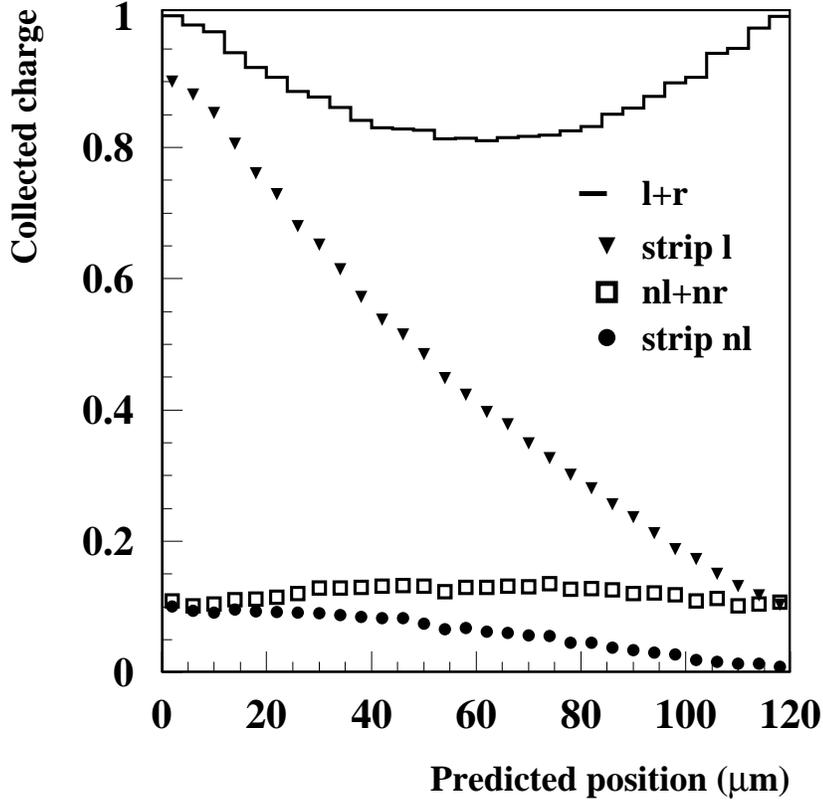,width=13cm}
\end{center}
\caption[Collected charge by the readout strips as a function of the 
predicted position]{\label{chardiv_plot}  Sum of the charge collected by the left 
and right  readout strips (sum l+r, solid line), the charge collected by the left readout strip (strip l, triangles) falling from the 
interstrip impact position  x=0 $\mu$m to  the interstrip impact position x=120 $\mu$m, charge collected  by the  next left  
neighbour of  the left readout strip (strip nl, dots) and sum of the charge collected by both the next-to-closest 
readout  strips (sum nl+nr, open squares). }
\end{figure}

  The charge collected 
 at the strips close to the particle incident position
 is shown in figure~\ref{chardiv_plot}. The two readout strips closest to the particle incident position 
 and the next neighbours are labeled strips l (left), r (right),   nl (next left) and nr (next right). 
 The most probable value of the charge 
 (obtained by fitting the distribution to a Landau curve) collected  by
 strip l alone and  by the sum of strips la nd r (l+r) are plotted in bins of the interstrip impact position predicted by the telescope.
 The  normalization factor of all distributions, denoted   S$^{\rm max}_{\rm l+r}$,  is determined  as  the sum of the signals  detected on strip l and r when a particle crosses  the detector exactly underneath a readout strip.
When a particle crosses the detector between two readout strips (l and r)
the charge collected on the intermediate strips induces charges on the
readout ones producing a dependence on  the distance
 to the impact position with large deviations from linearity near the readout strips.
Due to capacitative couplings between the strip implants,  the 
fraction of charge collected by strip r is not negligible 
($\sim10\%$ S$^{\rm max}_{\rm l+r}$) even for the case of particles crossing the detector exactly
underneath  the readout strip l (positions x=0 in figure \ref{chardiv_plot})
and vice versa. The charge collected by the next left neighbour of the readout strip l, nl, and 
the sum (nl+nr) of the signal collected by the two neighbours to the closest readout strips l (nl) and r (nr) is also shown in figure~\ref{chardiv_plot}.
The measurement demonstrates that a simple model considering only capacitances to
neighbouring strip implants resulting in charge collected only at  the two
neighbour strips is not satisfactory. 
 Moreover, the sum of the two signals collected by strip l and r is not completely flat  as a function of the 
interstrip position, showing a dip (-19~\% S$^{\rm max}_{\rm l+r}$) when the particles cross the detector in 
the central region between the two readout strips. This is mainly due to charge losses to the backplane. Taking 
into account the charge sharing to the next-to-closest readout strips, the effective charge loss to the  backplane 
is of the order of $\sim$ 16 \% .

\section{Simulation of the MVD detector response}
\label{sec:simulation}
 Diffusion, ionisation fluctuations, noise and 
 charge division were included in the simulation program which is
described in~\cite{bib:moritz_phd},
and  is based on \cite{bib:simulation}.
 Charge is generated inside the detector along the particle's path
 implementing ionisation fluctuations tuned to other measurements with silicon detectors~\cite{bib:simulation}. The charge drifts to the detector surface under the effect of the electric field;  
it is then assumed to be collected by  the closest strip implant~\cite{bib:simulation}.

Once the charges are collected on the strip implants they have
to be transferred to the readout strips. The capacitive network is more
complicated than the simple sketch in figure~\ref{fig:mvddet}, since also
capacitances to next to strip implants even further
apart are taken into account \cite{bib:jan}. Charge transfer coefficients
have been determined from testbeam measurements. They give the fraction of charge on a strip implant which is
transferred to the surrounding readout strips. To measure these coefficients only tracks crossing the detector 
 within 5\,$\mu$m underneath strip implants were used. 
 The strip implants are numbered from $\#1$ to $\#7$, starting from the 
 left readout strip from the impact position. 
The charge collected (i.e. the most probable value of the energy loss distribution) on the four surrounding 
readout strips, denoted as next left (nl), left (l), right  (r) 
and next right (nr),  is measured for tracks in positions $\#1$ to $\#7$.  
High statistics data samples have been used in order to achieve an accuracy  better than 1\%.
The collected charge reaches a maximum for positions  $\#1$ and $\#7$; 
 all coefficients are normalized to this value.  The detector response 
 is  assumed  to be symmetric. 
In the simulation all charges collected on a strip implant are
 transferred to the four surrounding readout strips using these measured
 coefficients (fractions smaller than 0.4$\%$ were neglected).
\begin{table}
\caption[Fraction of charge collected by four readout strips surrounding the 
particle position]{\label{tab:cdsim} Charge fraction collected by
 the four readout strips surrounding the particle impact position.
 Only particles crossing  the detector (between strips l and r) directly underneath strip 
 implants are selected. } 
\begin{center}
\begin{tabular}{l r r r r r } \hline
& \multicolumn{4}{l}{Charge transfer coefficient}& \\ \cline{2-5}
Particle Position{\hspace{4cm}}    & nl &   l    & r    & nr   & sum \\ \hline
$ \#1$: strip l & 0.091 &0.815 &0.091 & 0.004& 1.000\\ \hline
$ \#2$: 1$^{st}$ intermediate{   }& 0.082&0.655& 0.158&0.021&0.916\\ \hline
$ \#3$: 2$^{nd}$ intermediate{   }& 0.076&0.486& 0.256&0.041&0.859\\ \hline
$ \#4$: 3$^{rd}$ intermediate{   }& 0.058&0.363& 0.363&0.058&0.842\\ \hline
$ \#5$: 4$^{th}$ intermediate{   }& 0.041&0.256& 0.486&0.076&0.859\\ \hline
$ \#6$: 5$^{th}$ intermediate{   }& 0.021&0.158& 0.655&0.082&0.916\\ \hline
$ \#7$: strip r & 0.004&0.091& 0.815&0.091& 1.000\\ \hline
\end{tabular}
\end{center}
\end{table}
For every readout channel an additional signal according to Gaussian
distributed noise was simulated. The width of the Gaussian was chosen
in order to obtain the same S/N as measured in the testbeam data.
 
Comparison between the results of the simulation program with the data measurements
is presented in section~\ref{sec:comp_simul}.

\section{Position reconstruction for perpendicular tracks}
\label{sec:ene_reso}

\subsection{The eta algorithm.}
A standard method to reconstruct the impact position, proven to work at small incidence angle, 
is the so called $\eta$ algorithm \cite{bib:etaalg}, \cite{bib:etaalg1} . 
It consists of a non-linear interpolation between the two neighbouring strips of  
the cluster which have collected the highest signals (indicated in the following 
as S$_{\rm{right}}$ and S$_{\rm{left}}$, respectively). 
For each event, the quantity $\eta$:
\begin{equation}
  \eta = \frac{S_{\rm{right}}}{S_{\rm{right}}+S_{\rm{left}}}
\end{equation}
is calculated. 
\begin{figure}[h]
  \centerline{\epsfig{figure=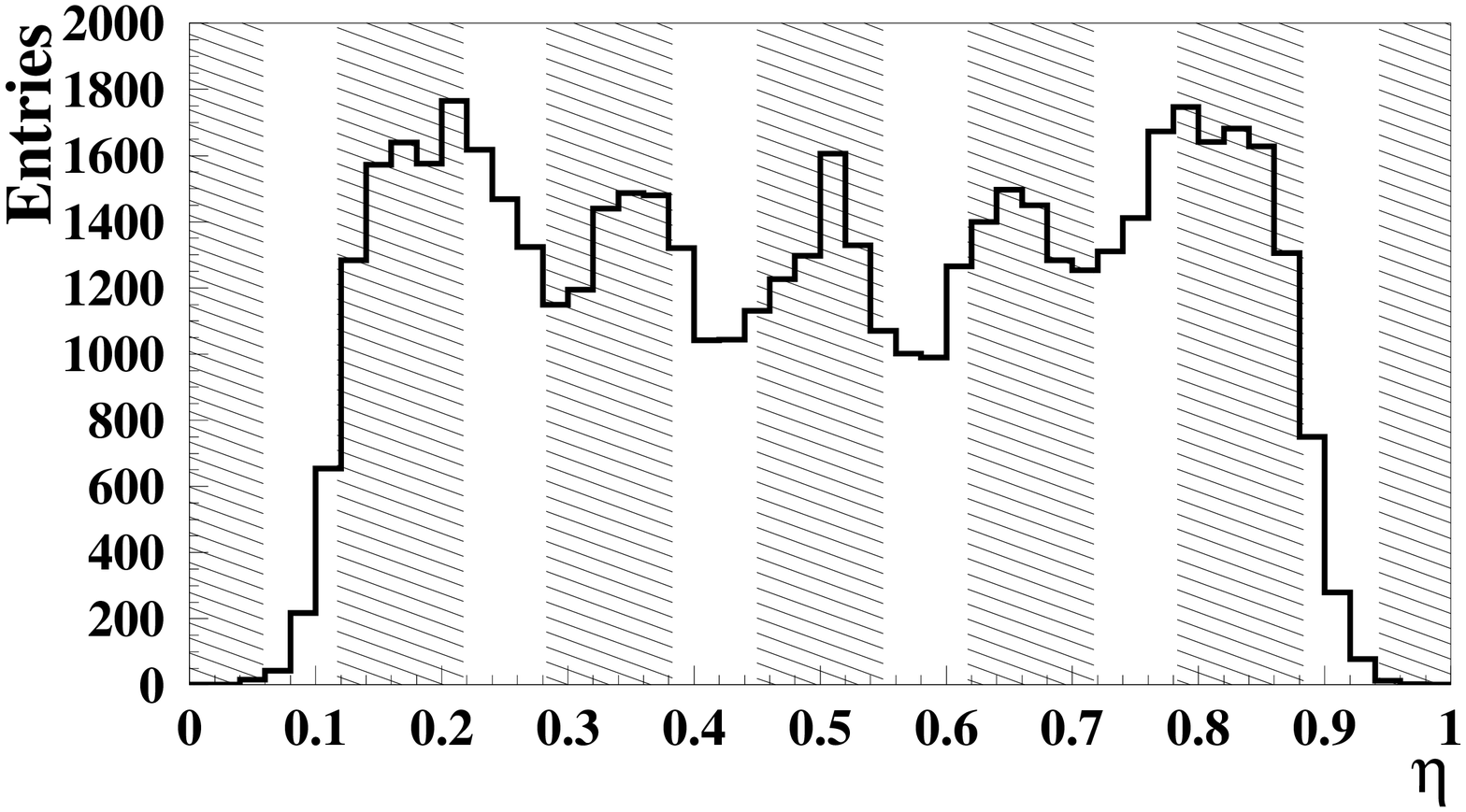,width=12cm}}
  \caption[Example of $dN/d\eta$ distribution for the DUT]
  {\rm Example of $dN/d\eta$ distribution.
 The dashed area covers the position of the p$^{+}$ implants
 (a  linear dependence of  $\eta$ on the interstrip hit position  is assumed). 
 } 
  \label{fig:eta}
\end{figure}
Figure \ref{fig:eta} shows the $dN/d\eta$ 
distribution for the MVD detector obtained from the  testbeam data: in the region close to the readout strips
there are very few entries, and clear peaks can be seen elsewhere.  
In case of a fully linear behaviour, the $\eta$ distribution would 
be completely flat and the dashed bands (where the peaks are observed) would represent the position of the intermediate strips. The fact that the $dN/d\eta$ is not uniform,
 although the beam profile is uniform over the detector area,
 indicates that the capacitive charge division mechanism  is not fully linear
 in the hit position between readout strips. To correct for this non-linearities a probability
density function (an example is shown in figure  ~\ref{fig:corr_func})
is determined from the data:
\begin{equation}
  f(\eta_0) = \frac{1}{N}\cdot \int_{0}^{\eta_0} \frac{dN}{d\eta} d\eta
\end{equation}    
where N is the total number of entries in the $dN/d\eta$ distribution and $\eta_0$
is the $\eta$ value for the considered event.
The corrected impact position is then given by:
\begin{equation}
  y_{\rm{rec}} = {\rm{p}}\cdot f(\eta_0)+ y_{\rm{left}}
\end{equation} 
where p is the readout pitch of 120 $\mu$m and $y_{\rm{left}}$ denotes the position of the 
left strip. 

\subsection{Intrinsic resolution}
\label{sec:intres}
The intrinsic resolution is defined as the spatial precision of the MVD 
 detector
\begin{figure}
\centerline{\epsfig{figure=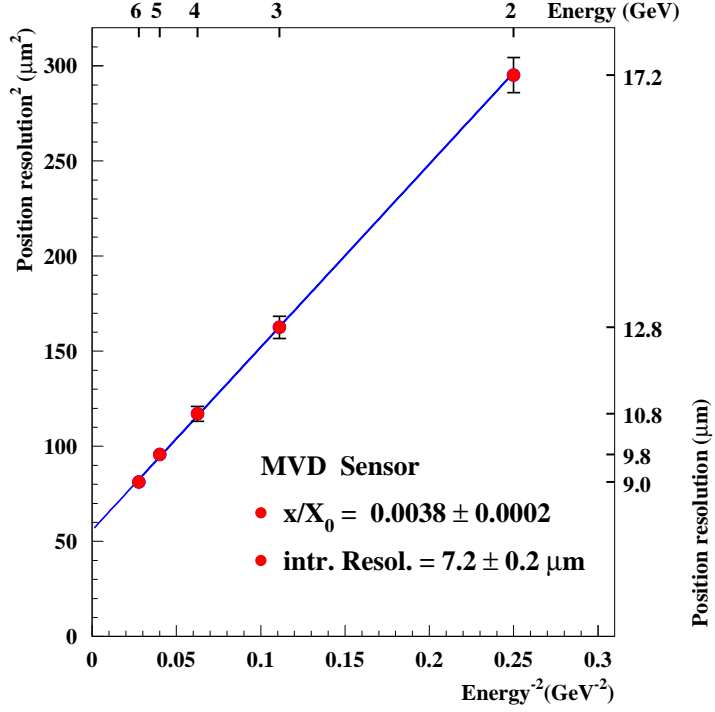,width=10cm}}
  \caption[Position resolution as a function of the beam energy]
  {\rm Position resolution as a function of the beam energy. The result of the fit for the ratio 
    $x/X_0$ (where x is the thickness of the material and $X_0$ the radiation length) is also shown.}
  \label{fig:resol_xc1}
\end{figure}
 
The measured position resolution $\sigma _{\rm res}$, defined as the width of the residual distribution obtained from 
a fit to a Gaussian function, includes several contributions:
\begin{equation}
  \label{eqn:res_tot}
  \sigma_{\rm res} = {\sqrt{ (\sigma _{\rm MVD}^{\rm intr})^2 + 
  k\cdot (\sigma _{\rm tele}^{\rm intr})^2 +
  \sum_{i} k_{i}\cdot \Delta \theta ^{2}_{\rm ms}}}
\end{equation}
where $\sigma _{\rm MVD}^{\rm intr}$ is the intrinsic resolution of the 
MVD detector, $\sigma _{\rm tele}^{\rm intr}$ is the intrinsic resolution of the telescope 
sensors, (k, k$_i$) are geometrical factors~\cite{bib:moritz_phd} related to the relative distances between the 
telescope modules, the MVD detector and also including the thickness of the aluminium window foils, and  
\mbox{$\sum_{i} k_{i}\cdot \Delta \theta_{\rm ms}$} is the extrapolation error due to the 
multiple Coulomb scattering along the particle direction,
 $\Delta \theta^{2}_{\rm ms} \propto p_{\rm beam}^{-2}$~\cite{bib:PDG}.
  
To extract the intrinsic position resolution of the MVD detector ($\sigma_{\rm MVD}^{\rm intr}$), 
the contributions of the  second (see subsection~\ref{ss:telescope}) and third term in equation  
\ref{eqn:res_tot} have been evaluated.  
The effect of the multiple Coulomb scattering at low beam energy (2-6 GeV)
cannot be neglected;
 the intrinsic resolution has been extracted by fitting the residual distribution 
measured at several beam energies (shown in  figure~\ref{fig:resol_xc1}) 
to the formula in  equation \ref{eqn:res_tot}.
Since the production of $\delta$-rays can spoil the position resolution,  (see subsection~\ref{ss:delta}),
 a selection cut to reject the events in the tail of the 
energy loss distribution has been applied in the previous calculation:
\[S_{\rm cluster} \leq 1.7 \cdot S_{\rm peak}\] 
where $S_{\rm cluster}$ is the total cluster charge and $S_{\rm peak}$ is the most probable 
energy  deposition. 
The intrinsic position resolution obtained for the MVD detector  at $\theta = 0^\circ$ incidence angle is:
\[\sigma_{\rm MVD}^{\rm intr} = 7.2 \pm 0.2 \;\mu {\rm m}\] 

 Different strip lengths do not affect significantly neither 
 the resolution nor the charge division mechanism~\cite{bib:redondo_phd}.

\subsection{Resolution vs interstrip hit position}
\begin{figure}[h]
\begin{center}
  \mbox{ 
\epsfig{figure=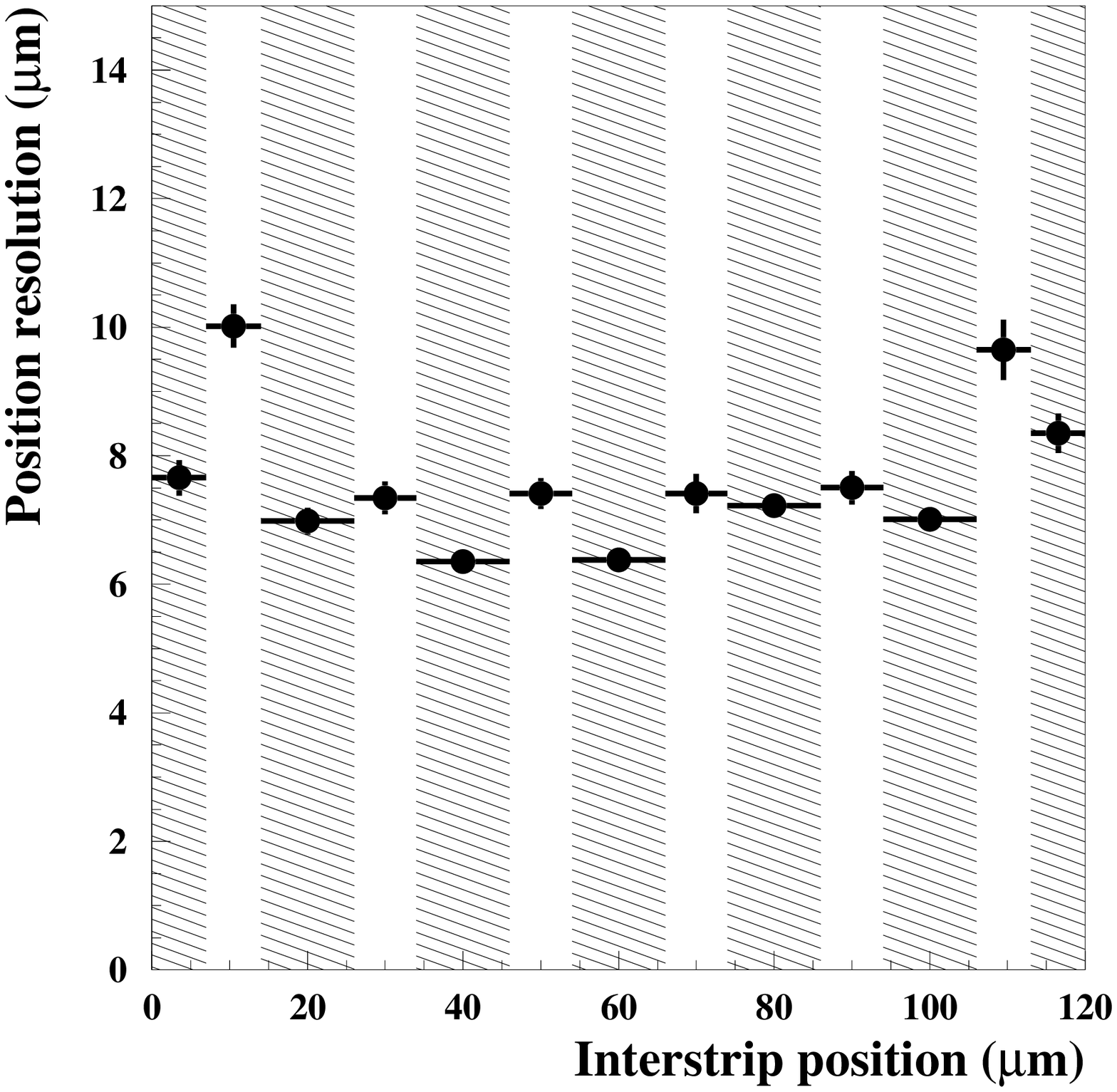,width=0.49\linewidth}
 \put(-1.9,1.8){(a)}
\hfill
\epsfig{figure=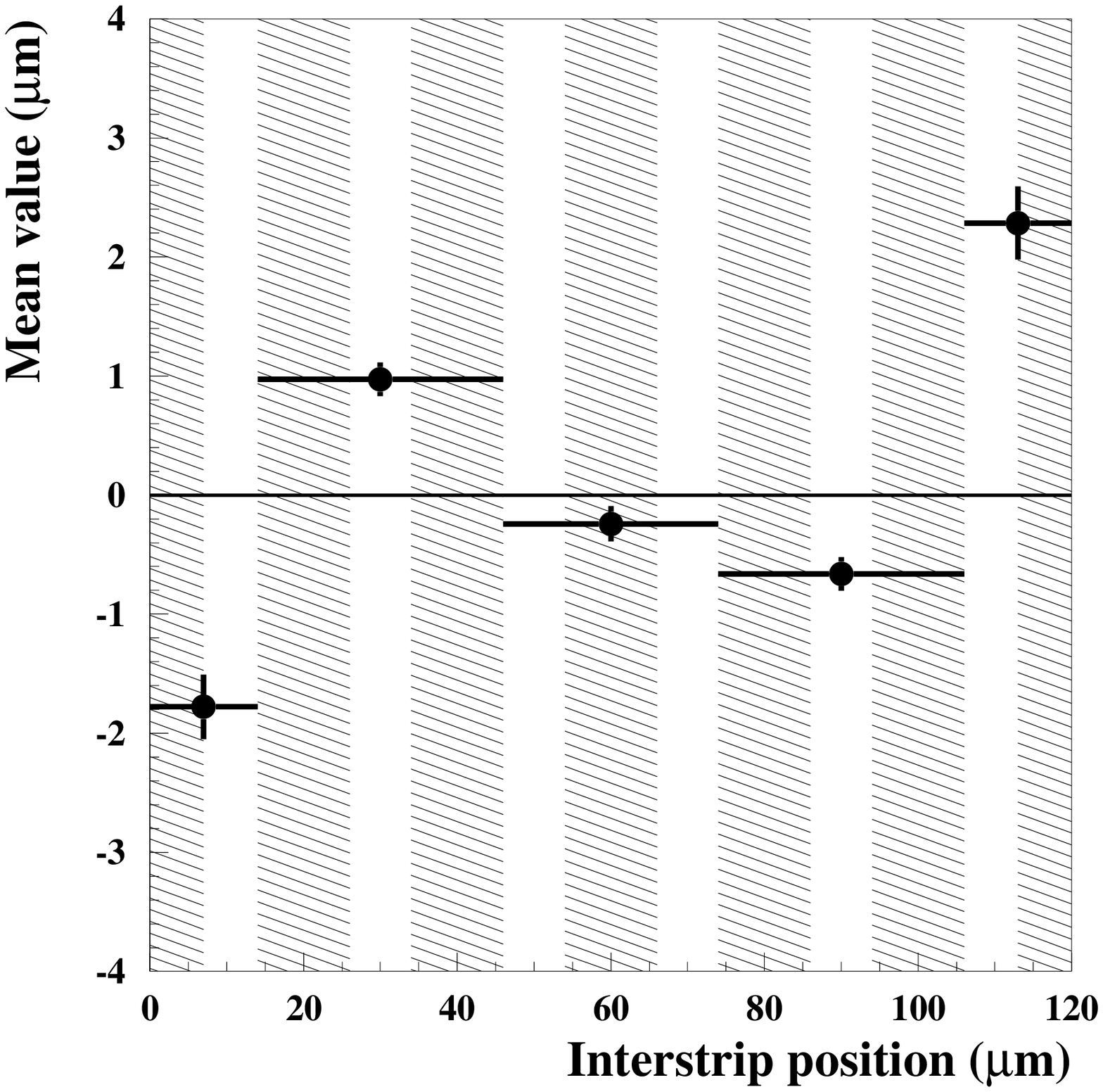,width=0.49\linewidth}}
 \put(-1.9,1.8){(b)}
  \caption[Position resolution and average residual shift from the origin as a function of
  the interstrip hit position]
  {\rm (a) Position resolution and (b) average residual shift from the origin as a function of
  the interstrip hit position. Note that in this case no cut to reject $\delta$-electrons
  has been applied. The hatched bands indicate the position of the readout and intermediate
  strips.}
  \label{fig:resandpeak0}
\end{center}
\end{figure}

Figure \ref{fig:resandpeak0}(a) shows the position resolution as a function of the 
interstrip hit position. 
 The presence of an alternate systematic 
pattern when moving  from a p$^+$ implant to the next one is noticeable. However, the variations observed
 are in general very 
small ($\lsim$ 1$\mu$m).  In the region close to the readout strips,
 the  position resolution becomes slightly worse. This effect is a consequence of the use of only 
two readout strips for the position reconstruction: when a particle traverses the silicon 
sensor very close to a readout strip, a high signal is induced on that readout strip and 
only relatively small signals on the neighbouring ones. Therefore the $\eta$ algorithm 
becomes more sensitive to noise fluctuations.
Figure~\ref{fig:resandpeak0}(b) shows the mean value of the Gaussian fit 
to the residual distribution (corresponding to a systematic shift from the origin) as a function 
of the interstrip hit position. Close to the readout strips the systematic shift in the position 
reconstruction is larger ($\sim$ 1.5-2 $\mu$m) than  in the central area between readout 
strips ($\lsim$ 1 $\mu$m). 

\section{Position reconstruction for small angle of incidence tracks}
\label{sec:smallangle}
\subsection{The 3-strips algorithm}
\begin{figure}[h]
  \centerline{\epsfig{figure=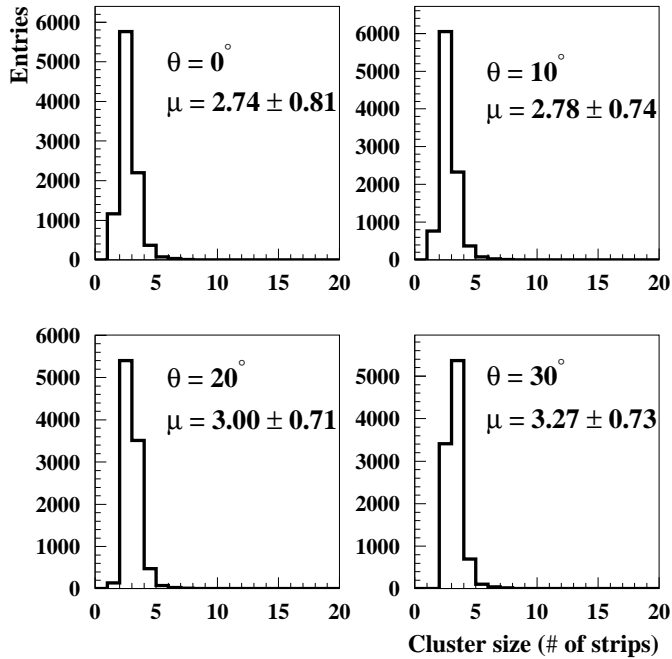,width=10cm}}
  \caption[Cluster size distribution for small incidence angles]
  {\rm Cluster size distribution for small incidence angles 
    (0$^ \circ$, 10$^ \circ$, 20$^ \circ$, 30$^ \circ$).
The mean value and the rms  of the cluster size distribution are also indicated.}
  \label{fig:clus0}
\end{figure}

The use of only two readout strips to extract position information may not be the most appropriate
choice for the MVD detectors, since the symmetric charge sharing to the next-to-closest
readout strips is not negligible. Figure \ref{fig:clus0} shows the cluster 
size distribution at four different incidence angles (0$^ \circ$, 10$^ \circ$, 20$^ \circ$, 30$^ \circ$). 
At $\theta =$ 0$^ \circ$, already  $\sim$ 28\% of all cluster consists of more than two strips 
and this percentage becomes much larger ($\gsim$ 45\%) for $\theta \gsim $20$^ \circ$. 
An alternative  position reconstruction algorithm, which makes use of three strips,  
has been therefore developed. 
From each cluster, the strip with the highest signal, k,  and its closest 
neighbours (k-1,k+1)  are selected  and the following quantities are calculated:
\begin{equation}
  {\rm p_ {left}} = \frac{S_{k}\cdot k + S_{k-1}\cdot (k-1)}{S_{k}+S_{k-1}} \;\;\; {\rm and} \;\;\;
  {\rm p_ {right}} = \frac{S_{k}\cdot k + S_{k+1}\cdot (k+1)}{S_{k}+S_{k+1}}
\end{equation}
The uncorrected reconstructed position p$_{\rm rec}$ (in analogy to the linear $\eta$ interpolation) 
is then defined as:
\begin{equation}
  {\rm p_{rec}} = \frac{{\rm p_{left}}\cdot w^{n}+{\rm
p_{right}}}{1+w^{n}} \;\;\;
  {\rm where} \;\;\; w = S_{k-1}/S_{k+1} \;;\;\; n=2.
\end{equation}
 $n=2$ is found to work better than $n=1$ because in the former
  case noise is suppressed by giving less weigh to the strip
  with the lowest charge. The
corresponding interstrip position $\tilde{p}$ is given by:
\[ \tilde{p} = {\rm mod}({\rm p_{rec}},1.) \]

\begin{figure}
  \centerline{\epsfig{figure=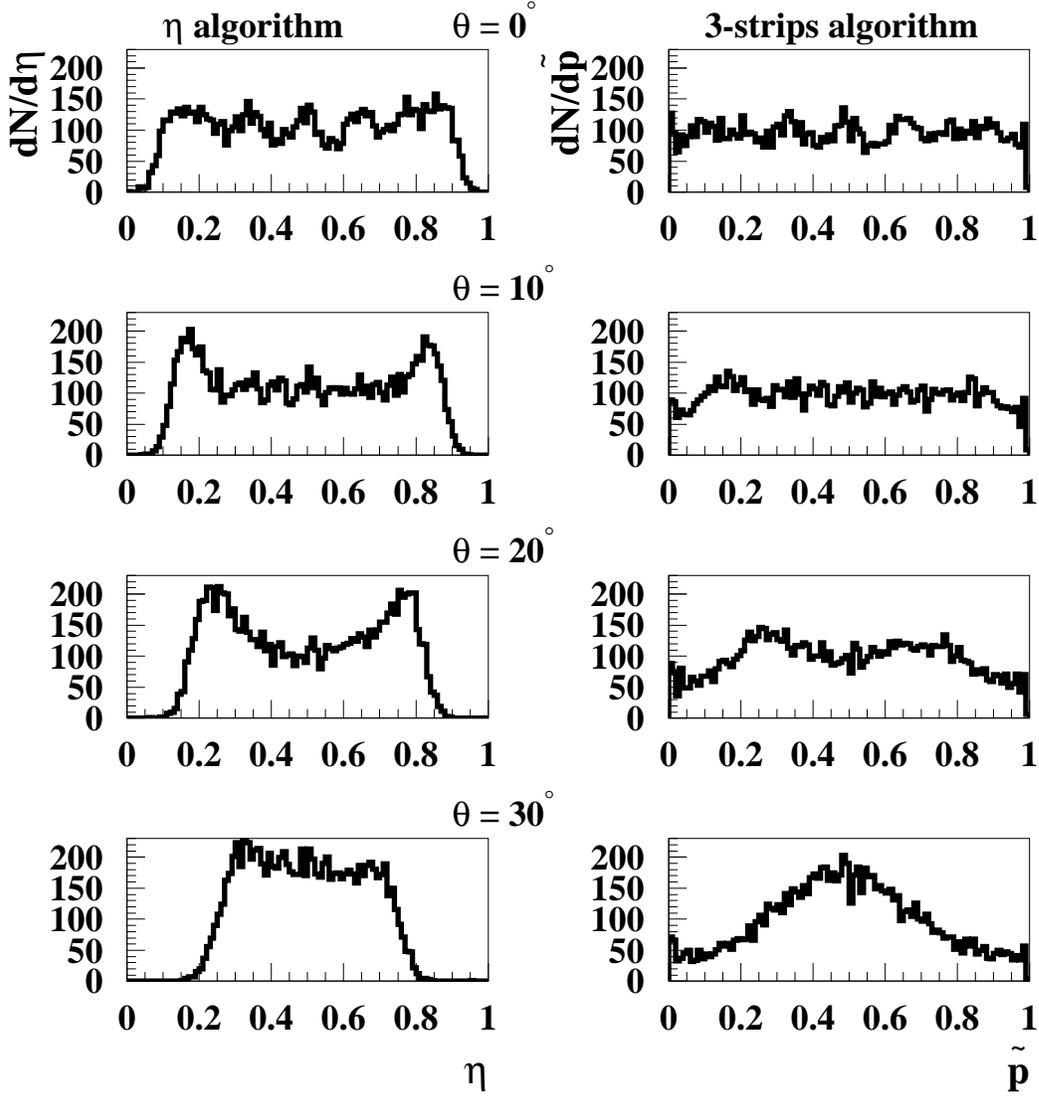,width=16cm}}
  \caption[dN/d$\eta$ and dN/d$\tilde{p}$ distributions for small incidence angles]
  {\rm dN/d$\eta$ and dN/d$\tilde{p}$ distributions for small incidence angles.}
  \label{fig:pos_comp}
\end{figure}
\begin{figure}
  \centerline{\epsfig{figure=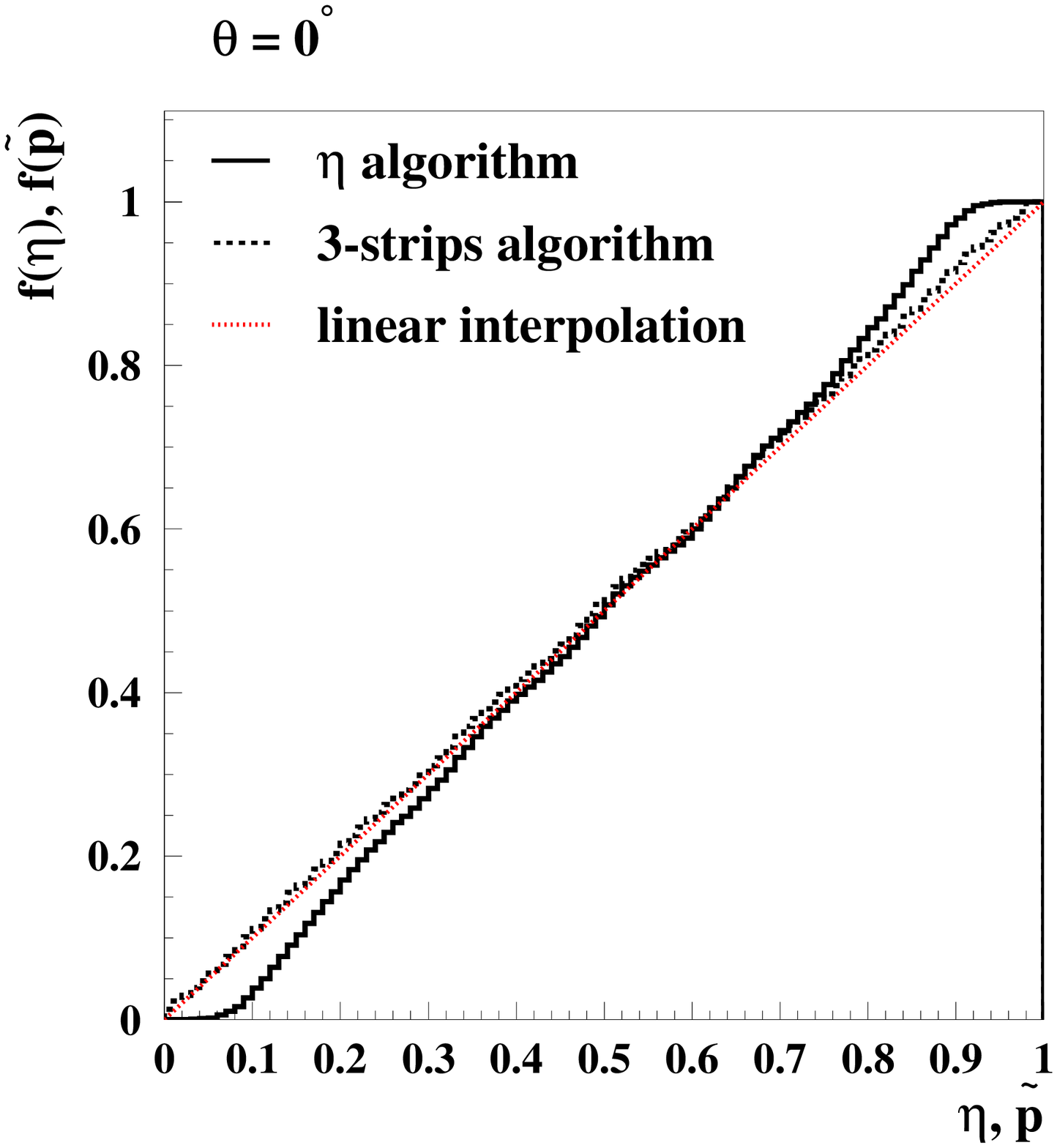,width=8.0cm}}
  \caption[Probability density function for the $\eta$ and the 3-strips algorithm]
  {\rm Probability density function for the $\eta$ and the 3-strips algorithm. Note that
    at $\theta = 0^{\circ}$ the result for f(${\tilde p}$) is very close to the linear interpolation.}
  \label{fig:corr_func}
\end{figure}

\begin{figure}
  \mbox{\hspace*{-1.6cm}
    \epsfig{figure=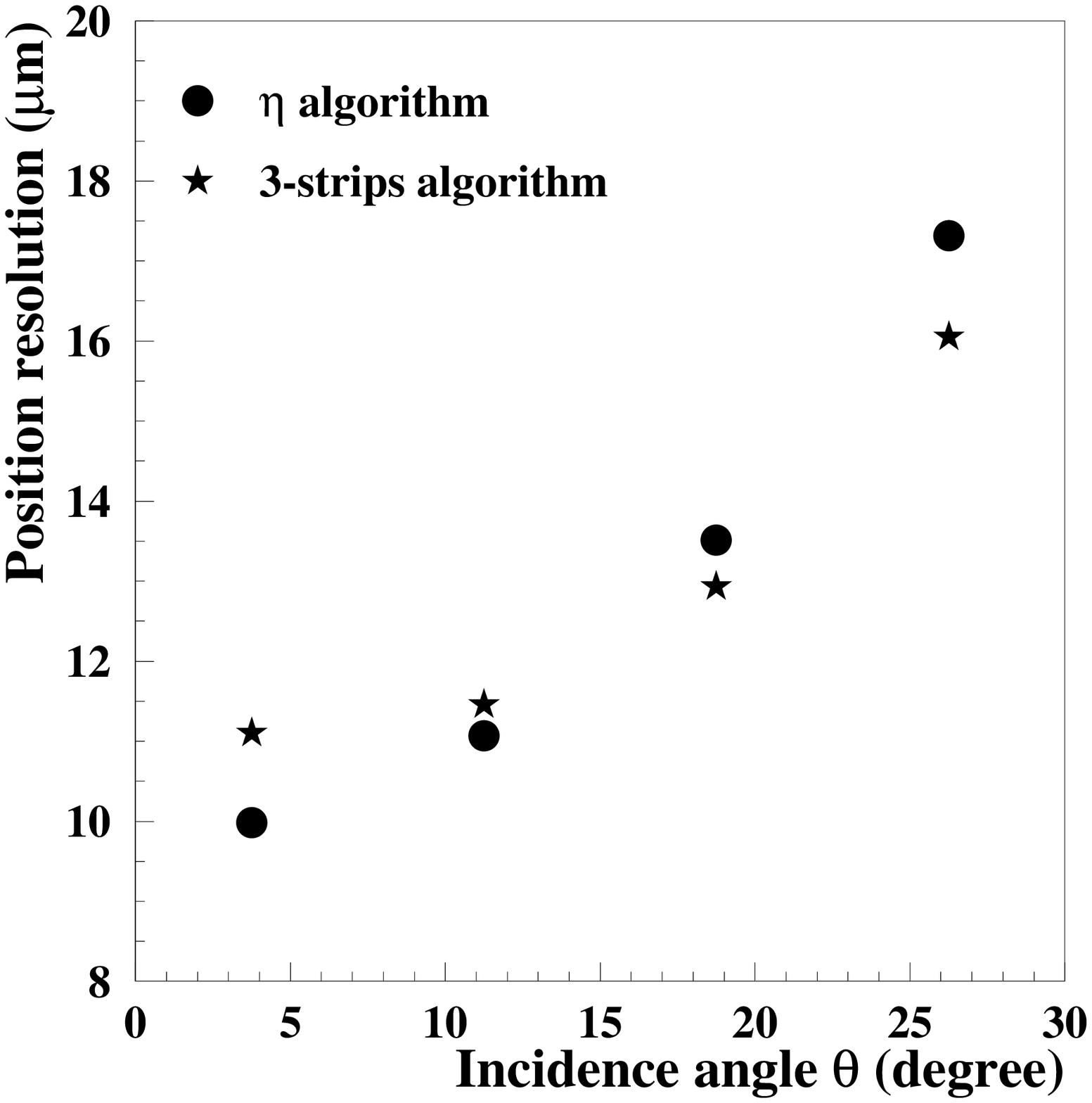,width=8.5cm}
    \put(-1.9,1.5){(a)}
\hspace{-0.6cm} 
    \epsfig{figure=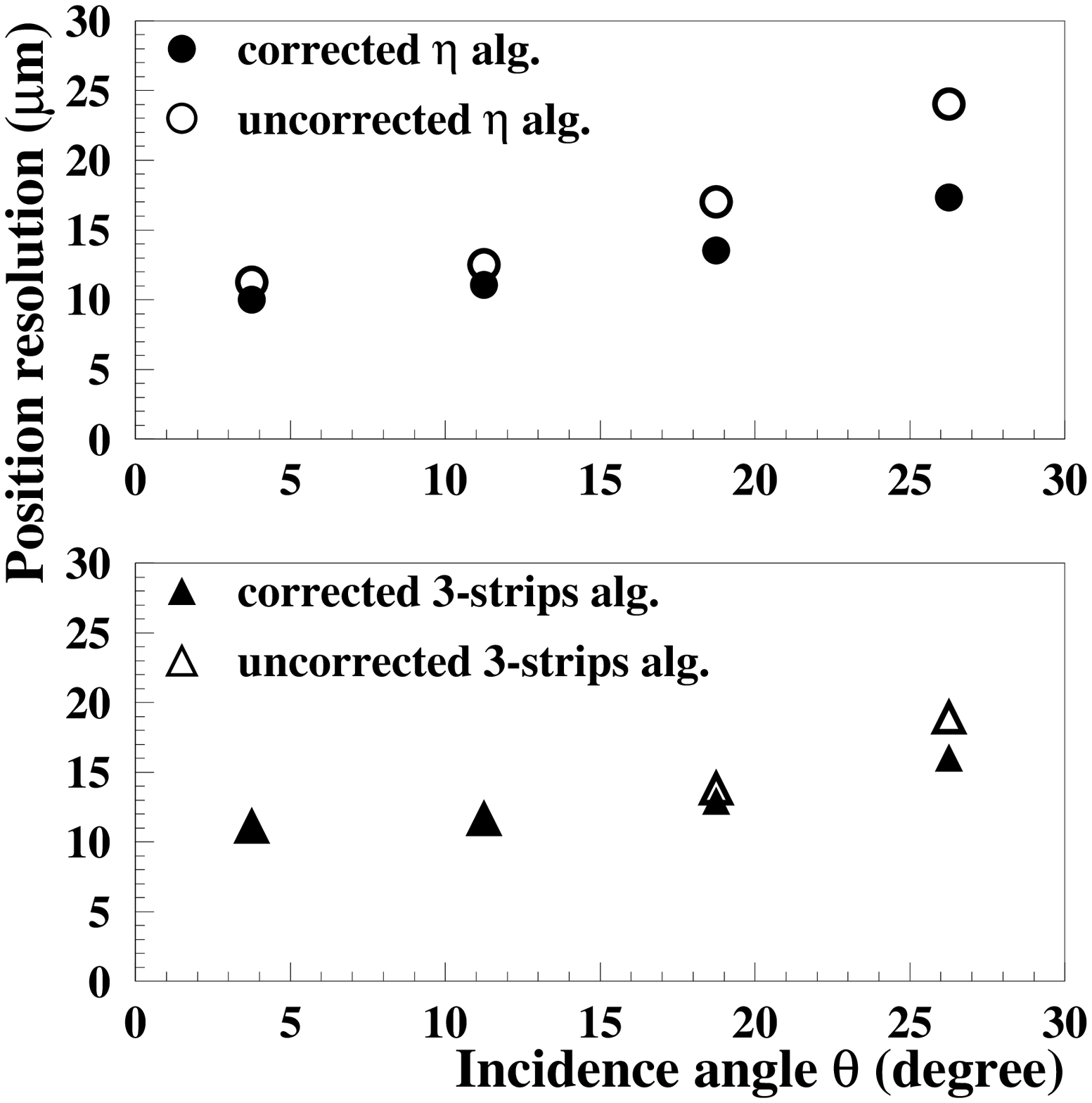,width=8.5cm}}
    \put(-1.9,1.5){(b)}
  \caption[Corrected and uncorrected position resolution for small incidence angles]
  {\rm Position resolution for small incidence angles 
    (0$^ \circ$, 10$^ \circ$, 20$^ \circ$, 30$^ \circ$). (a) Comparison 
    between the corrected position resolution calculated using the $\eta$ and the 3-strips 
    algorithm; (b) comparison between the uncorrected and corrected position 
    resolution for both algorithms.}
  \label{fig:res0_30}
\end{figure}

 
Figure \ref{fig:pos_comp} shows the dN/d$\eta$ and dN/d$\tilde{p}$ distributions for small 
incidence angles (0$^ \circ$, 10$^ \circ$, 20$^ \circ$, 30$^ \circ$). By means of the 3-strips algorithm
a more uniform distribution can be obtained. In analogy with the $\eta$ algorithm a probability
density function can be defined and 
the corrected impact position is given by:
\begin{equation}
  y_{\rm{rec}} = {\rm{p}}\cdot f(\tilde{p}_0)+ {\rm{p}}\cdot p_{\rm{left}}
\end{equation}
where p is the readout pitch and $p_{\rm{left}}$ denotes the position reconstructed using the left and the
central strips.
Figure \ref{fig:corr_func} shows the comparison between 
$f(\tilde{p})$ and $f(\eta)$ for  $\theta = {0}^\circ$.
The position resolutions obtained using the $\eta$ algorithm and the 3-strips algorithm 
for the four different incidence  angles (0$^ \circ$, 10$^ \circ$, 20$^ \circ$, 30$^ \circ$) are shown 
in figure \ref{fig:res0_30}.
The resolution at 0$^ \circ$ calculated with the 3-strips algorithm is slightly
worse than the one obtained with the $\eta$ algorithm.  
Nevertheless the non-linearity correction for the 3-strips algorithm is much smaller than the one needed 
for the $\eta$ algorithm up to 30$^\circ$, as shown in the right plot of figure \ref{fig:res0_30}.
Thus the 3-strips algorithm is much less sensitive to possible variations in the 
probability density function  due to  non-uniformities of the MVD detectors. 
A simple centre of gravity algorithm has been discarded since its use results in a significantly worse 
position resolution for particle crossing the detector in the central region between two readout strips and in addition 
it introduces sizeable systematic shifts in the position reconstruction.
\subsection{Performance as a function of the $\phi$ angle} 
\label{ss:phi} 

In the barrel section of the MVD the angle $\phi$ can be as large as 70$^{\circ}$.
However the  size of the optical bench in the testbeam setup limited the measurement 
to a maximum angle $\phi$ of 30$^{\circ}$.
In the forward section of the MVD the $\phi$ range is bigger than the 
$\theta$ range due to the orientation  and position  of the FMVD
strips with respect to the interaction point\footnote{This is true if
the curvature induced by the magnetic field is not taken into account.
For low momenta particles the angle $\theta$ can be also large.}. 
Although the maximum angle $\phi$ in the FMVD is only $\sim$30$^{\circ}$, 
this is  correlated  with physical quantities of interest such as the
pseudorapidity of the track; any systematic effect  could be thus relevant for
high momentum tracks as the one of the scattered electron at high $Q^2$.
In the data analysis for the FMVD sensors, the width of the residual distribution was
$\sim$2-3 $\mu$m larger than in the standard setup (used for BMVD sensors) because 
of the different  geometrical constraints (i.e. the telescope modules had to be moved away from the 
MVD sensor).
The expected increase of the signal  with the path length
inside the detector ($\propto cos(\phi)^{-1}$) was  observed~\cite{bib:redondo_phd}. 
\begin{figure}[hbtp]
\begin{center}
\centerline{\mbox{
\hspace*{-1.0cm}
\epsfig{file=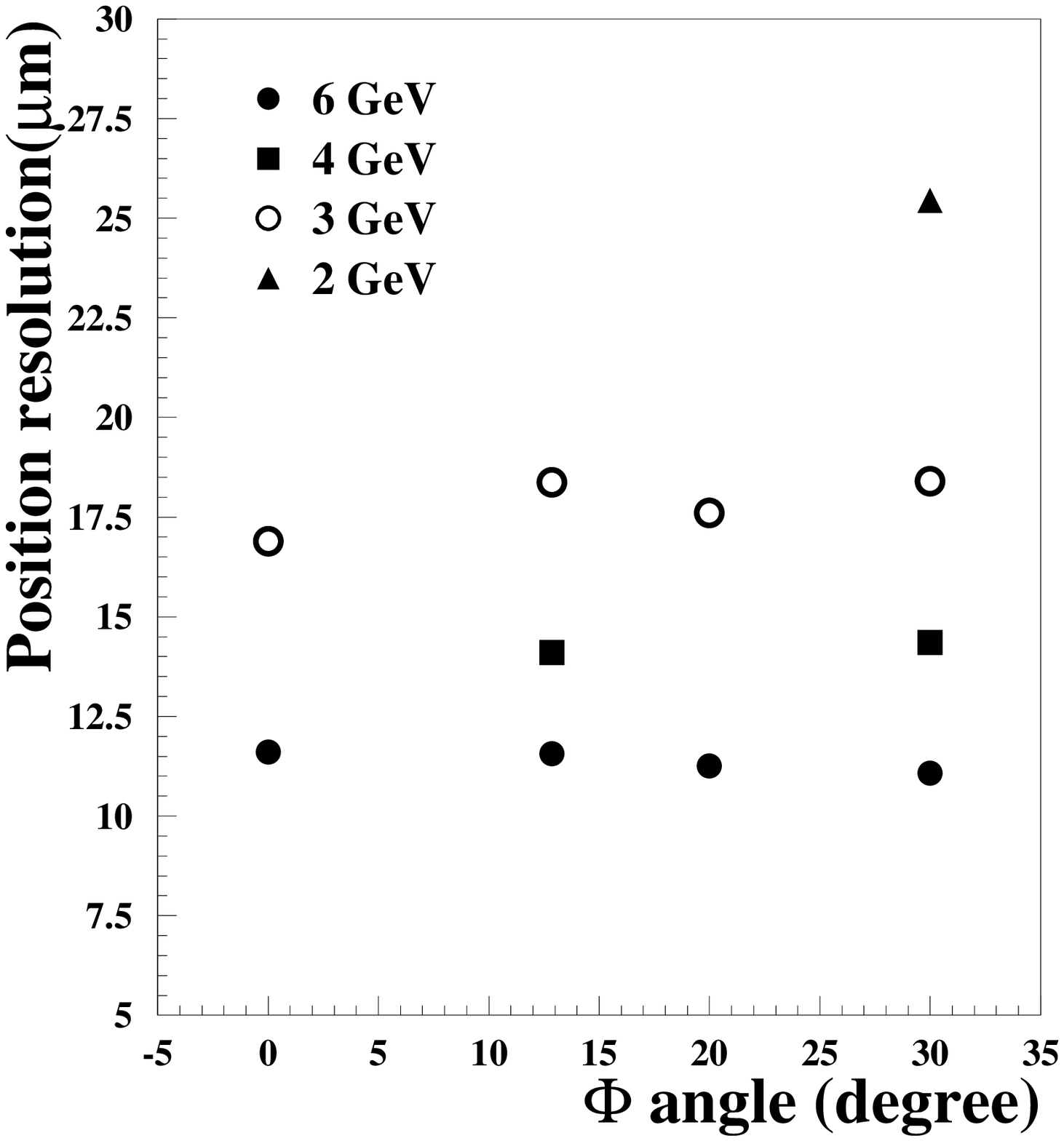,width=0.55\linewidth}
   \put(-20.3,1.8){(a)}
\epsfig{file=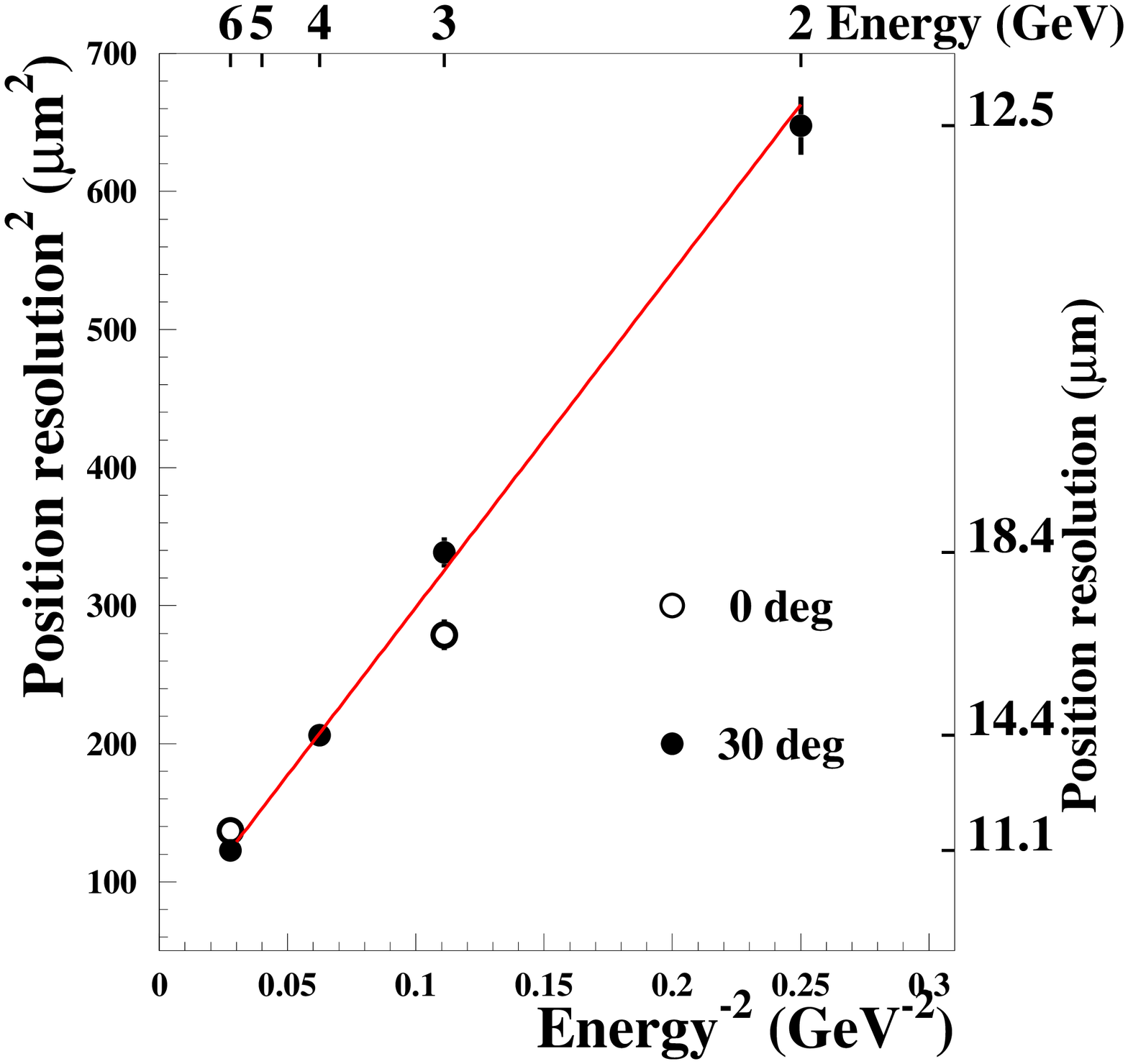,width=0.55\linewidth}
   \put(-20.3,1.8){(b)}
}}
\end{center}
\caption[$\sigma$ vs $\phi$]{(a) Position resolution as a function of  the angle $\phi$; (b)
$\sigma^{2}$ as a function of $1/E^2$. A fit to the data at $\phi=30^{\circ}$ is also shown.}
\label{f:the2}
\end{figure}

Figure ~\ref{f:the2}(a)  shows the position resolution as a function of the angle 
 $\phi$  for several beam energies. A rather flat behaviour is observed in the relevant  range for 
the FMVD. In figure~\ref{f:the2}(b) the squared position resolution as a function of $1/E^2$ is shown 
for two different angles (0$^{\circ}$ and 30$^{\circ}$). The contribution of multiple scattering shows up 
as a linear behaviour and the intrinsic resolution corresponds to the intercept at the origin (i.e. for 
infinite momentum particles) as it was discussed in subsection~\ref{sec:intres}. The data 
at  $\phi=30^{\circ}$ (squares) seem to  have a  larger slope and a smaller intercept compared with  
the results for $\phi=0^{\circ}$ (dots).  The larger slope can be attributed to the increase in the material 
of a factor $cos(30^{\circ})^{-1}\sim 1.15$.  In addition, the larger S/N obtained at larger $\phi$ angles can produce a better 
intrinsic resolution (as indicated by  the smaller intercept). 
 Systematic effects have been evaluated to be  smaller than 4\%,
   dominating  over the  statistical accuracy of $\sim$1.4\%.

\section{Position resolution for large angle of incidence tracks}
~\label{sec:ang_reso}

\begin{figure}
  \mbox{ 
    \hspace*{-1.cm}
\epsfig{figure=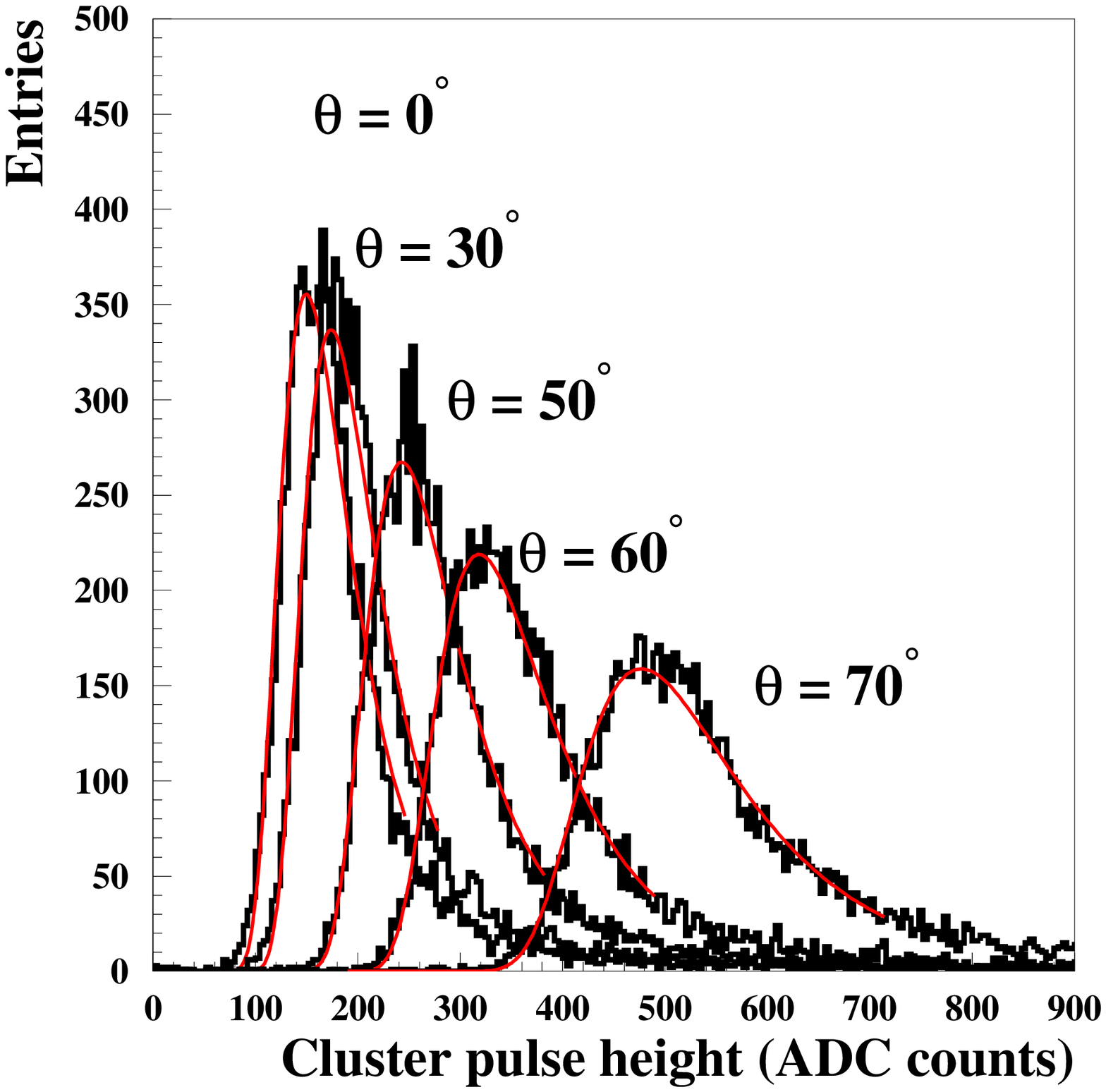,width=0.55\linewidth}
\put(-20,1.8){(a)}
    \hfill 
    \epsfig{figure=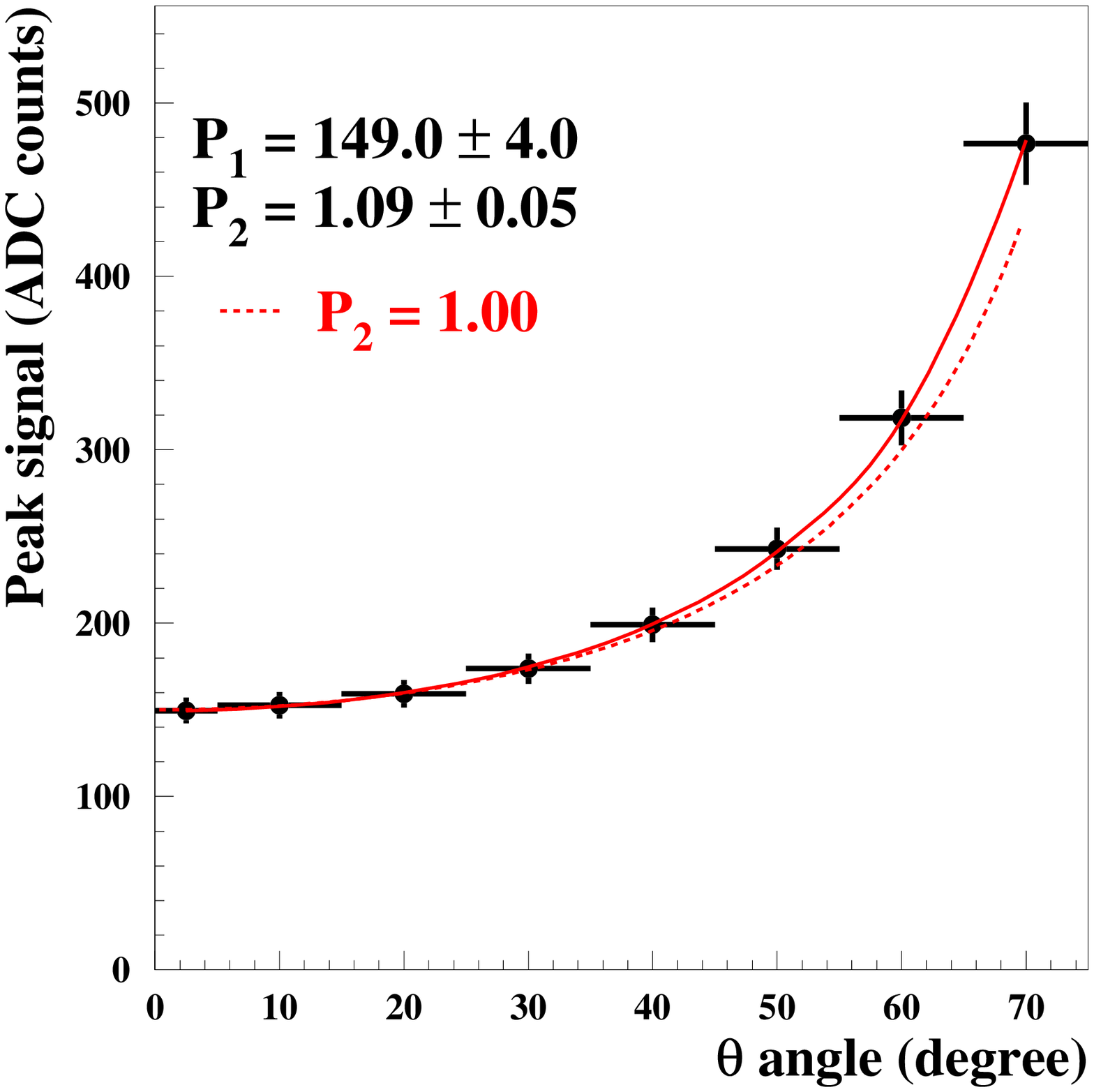,width=0.55\linewidth}}
\put(-20,1.8){(b)}
  \caption[Energy loss distribution at various incidence angles  and most probable
  value for the energy deposition as a function of the incidence angle]
  {\rm (a) Energy loss distribution at various incidence angles; (b) most probable
  value for the energy deposition as a function of the incidence angle.}
  \label{fig:landau_ang}
\end{figure}

At large incidence angles ($\theta > 30^\circ$) the charge is spread over several strips and 
the total cluster signal becomes larger as the particle's path length increases:
\begin{equation}
  \label{eqn:growth}
  S(\theta) \propto \frac{S(\theta = 0^\circ)}{\cos\theta}
\end{equation}
Since the central strips have on average the same signal, the information on the impact position is 
essentially contained only in the positions and signals of the cluster edge strips. Therefore 
reconstruction algorithms such as the $\eta$ one are inadequate to calculate the 
particle impact point on the detector.
\begin{figure}
  \centerline{
    \epsfig{figure= 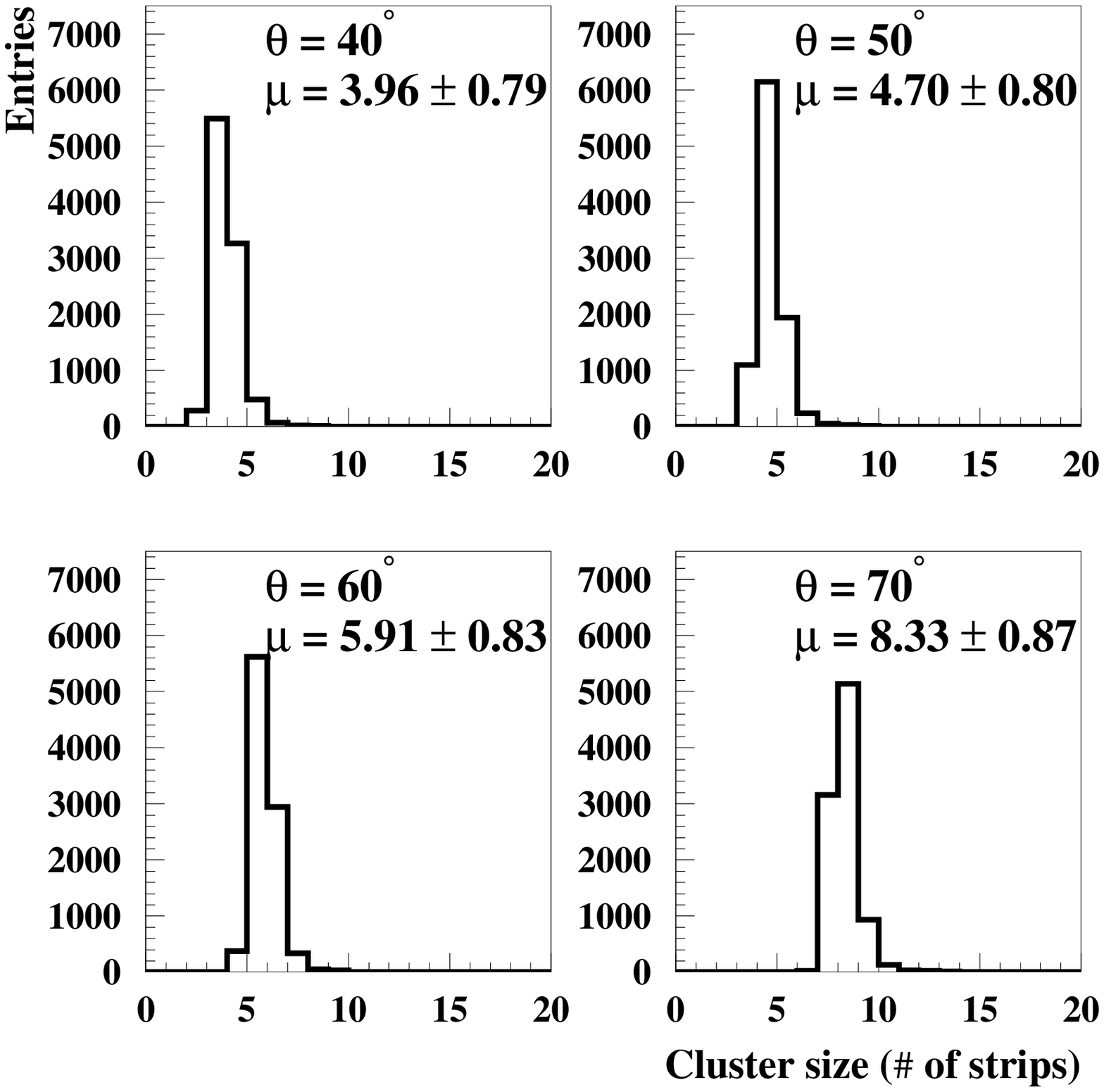,width=10cm}}
  \caption[Cluster size distribution for large incidence angles]
  {\rm Cluster size distributions for large incidence angles 
    (40$^ \circ$, 50$^ \circ$, 60$^ \circ$, 70$^ \circ$)}
  \label{fig:clus1}
\end{figure} 
Figure \ref{fig:clus1} shows the cluster size distributions for large angles of incidence   
(40$^ \circ$, 50$^ \circ$, 60$^ \circ$, 70$^ \circ$). 
Figure \ref{fig:landau_ang}(a) shows the energy loss distribution for various  
incidence angles ($ 0^\circ,\;30^\circ,\;50^\circ,\;60^\circ,\;70^\circ$), 
whereas the most probable value for the energy deposition as a function of the incidence angle is 
shown in figure~\ref{fig:landau_ang}(b). In this case the energy
deposition is defined by summing up the signals measured on $\pm$ 5 strips around the predicted
position:
\[S_{\rm tot} = \sum _{i = -5}^{+5} S_{i}\]
The result of a fit to the function:
\begin{equation}
  f(\theta,P_1,P_2) =  \frac{P_1}{\cos \theta^{P_2}}
\end{equation}
is also presented. The value $P_2 = 1.09 \pm 0.05$ is in agreement with the expectation 
in equation \ref{eqn:growth} (i.e. $P_2 = 1.0$).
\subsection{The head-tail algorithm}
A standard position reconstruction algorithm, proven to work at large incidence angle, 
is the so called `head-tail' algorithm  \cite{bib:etaalg1,bib:headtail}. All the strips with:
\[S_{\rm strip} > 3\cdot \sigma_{\rm chip}\]
where $S_{\rm strip}$ is the strip signal and $ \sigma_{\rm chip}$ is the average chip noise,
are considered. The first (head) and the last (tail) strips belonging to a cluster are selected  
and the impact position is defined as:
\begin{equation}
  y_{\rm rec} = \frac{y_{\rm head}+y_{\rm tail}}{2} + {\rm ht_{corr}} 
  \;\;\; {\rm with} \;\;\; {\rm ht_{corr}} = \frac{S_{\rm tail}-S_{\rm head}}{2\cdot <S>_{\rm strips}}\cdot p
\end{equation}
where $y_{\rm head}$ ($y_{\rm tail}$) is the position of the head (tail) strip, $S_{\rm head}$ 
($S_{\rm tail}$) is the corresponding signal, p is the strip pitch and $<S>$ is the average strip signal over the cluster.

The difference ($S_{\rm tail}-S_{\rm head}$) in the correction factor ${\rm ht_{corr}}$ is used to shift 
the average position $(y_{\rm head}+y_{\rm tail})/2$ towards the tail 
($S_{\rm tail}-S_{\rm head}>0$) or head ($S_{\rm tail}-S_{\rm head}<0$) strip of the cluster, 
taking into account the rough  proportionality of the energy loss to the particle 
path  in the detector.
\subsection{Comparison with the simulation}
\label{sec:comp_simul}
\begin{figure}
\centerline{\epsfig{figure=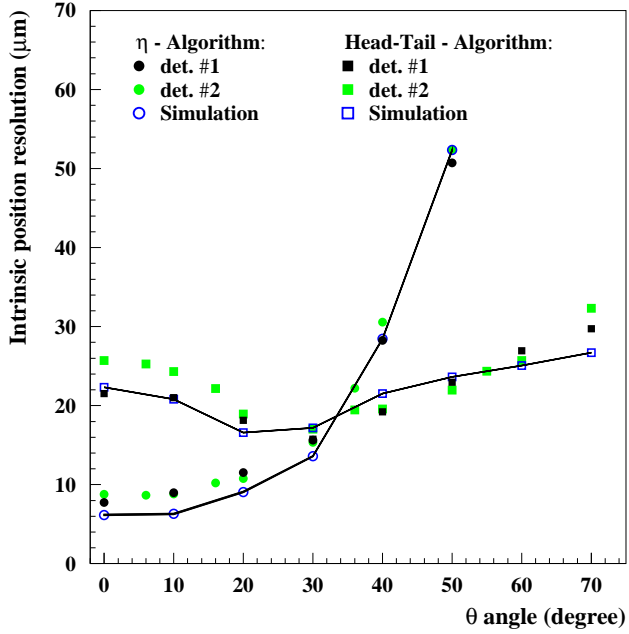,width=9.5cm}}
  \caption[Comparison between the $\eta$ and head-tail algorithms]
  {\rm Intrinsic position resolution calculated with the $\eta$ and head-tail algorithms
    as a function of the angle of incidence $\theta$.
}
  \label{fig:comparison}
\end{figure}

 Figure~\ref{fig:comparison} shows the position resolution calculated with the $\eta$ and head-tail algorithms as a function of the angle of incidence $\theta$.
    At angles $\lsim$ 30$^\circ$ the $\eta$ algorithm gives a much better resolution than 
     the head-tail algorithm. However for angles $\gsim$ 30-40$^\circ$, the latter proves to work 
     much better. The results using two different MVD barrel detectors  are presented and compared 
    with  the results obtained using the simulation program.
 The simulation is able to describe
the data over the whole angular region and for both reconstruction algorithms.  The S/N 
and charge transfer coefficients
of det. $\#1$ were used in the
simulation, which explains the slightly better agreement with this
detector. Since $\delta$-rays (see next section) are not taken into
account the simulated intrinsic resolution is found to be better than
the one obtained from testbeam data, especially
 for the $\eta$ algorithm
 at small incident angles. 
 The probability that the two highest strips in the
cluster, i.e. those  used by the $\eta$ algorithm,  are affected by a
$\delta$-ray which departs from the initial trajectory is smaller at
larger angles because the energy is  deposited along more strips.
\subsection{Optimisation of the reconstruction algorithm}
Reconstruction algorithms  such as the $\eta$, 3-strips  or the head-tail ones  are based on the
knowledge of the incidence angle $\theta$ of the particle, used to select
the most effective algorithm to be applied. 
Switching between different reconstruction algorithms clearly implies complications in the track reconstruction procedure.
Ideally an optimal algorithm should calculate impact positions with a single calculation 
procedure, using as little  additional external information  as possible to minimise systematic 
effects. 
 
\begin{figure}
  \mbox{
    \hspace*{-1.6cm}\epsfig{figure=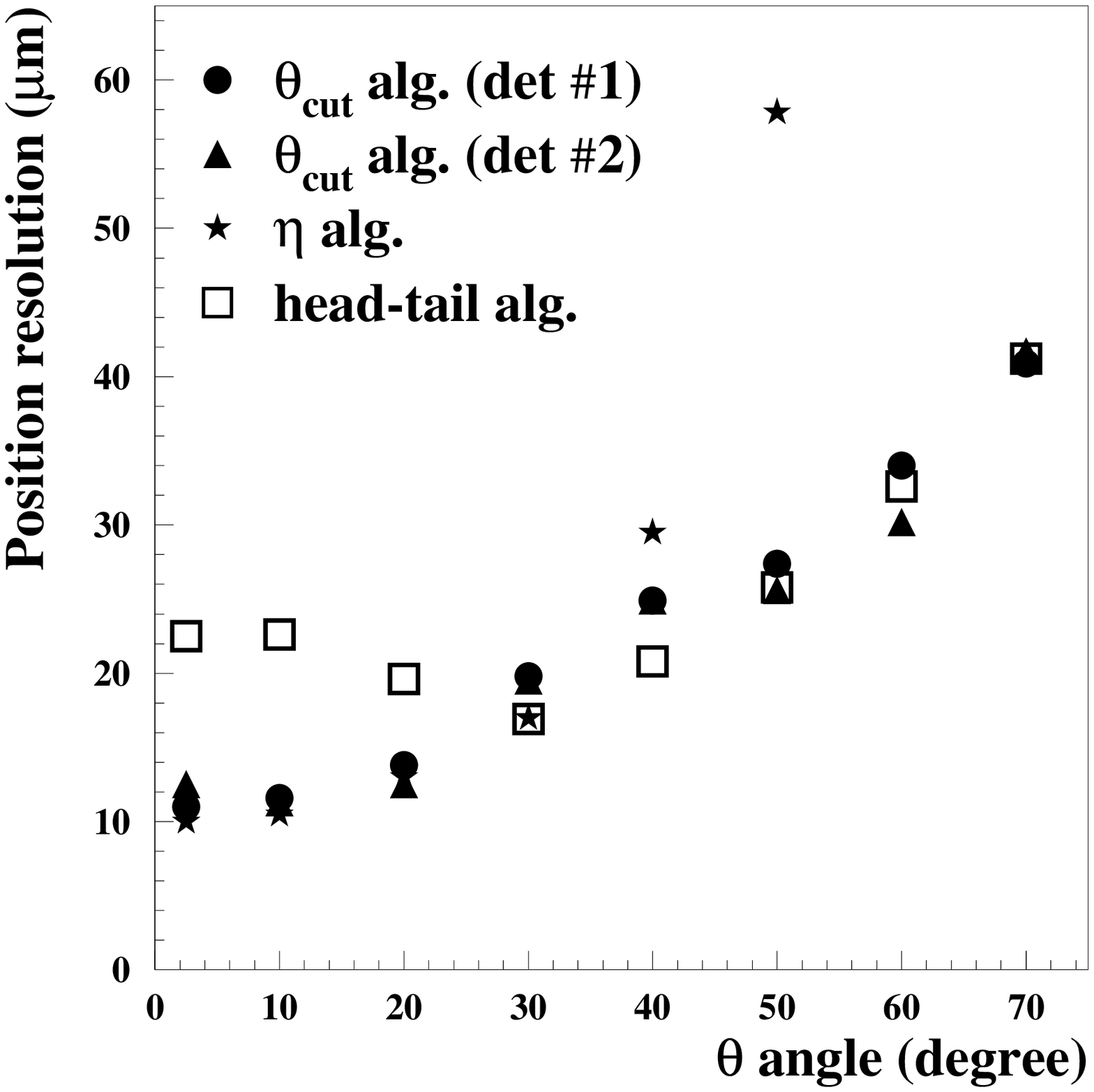,width=8.5cm}
 \put(-15,1.8){(a)}
    \hspace*{-0.8cm}\epsfig{figure=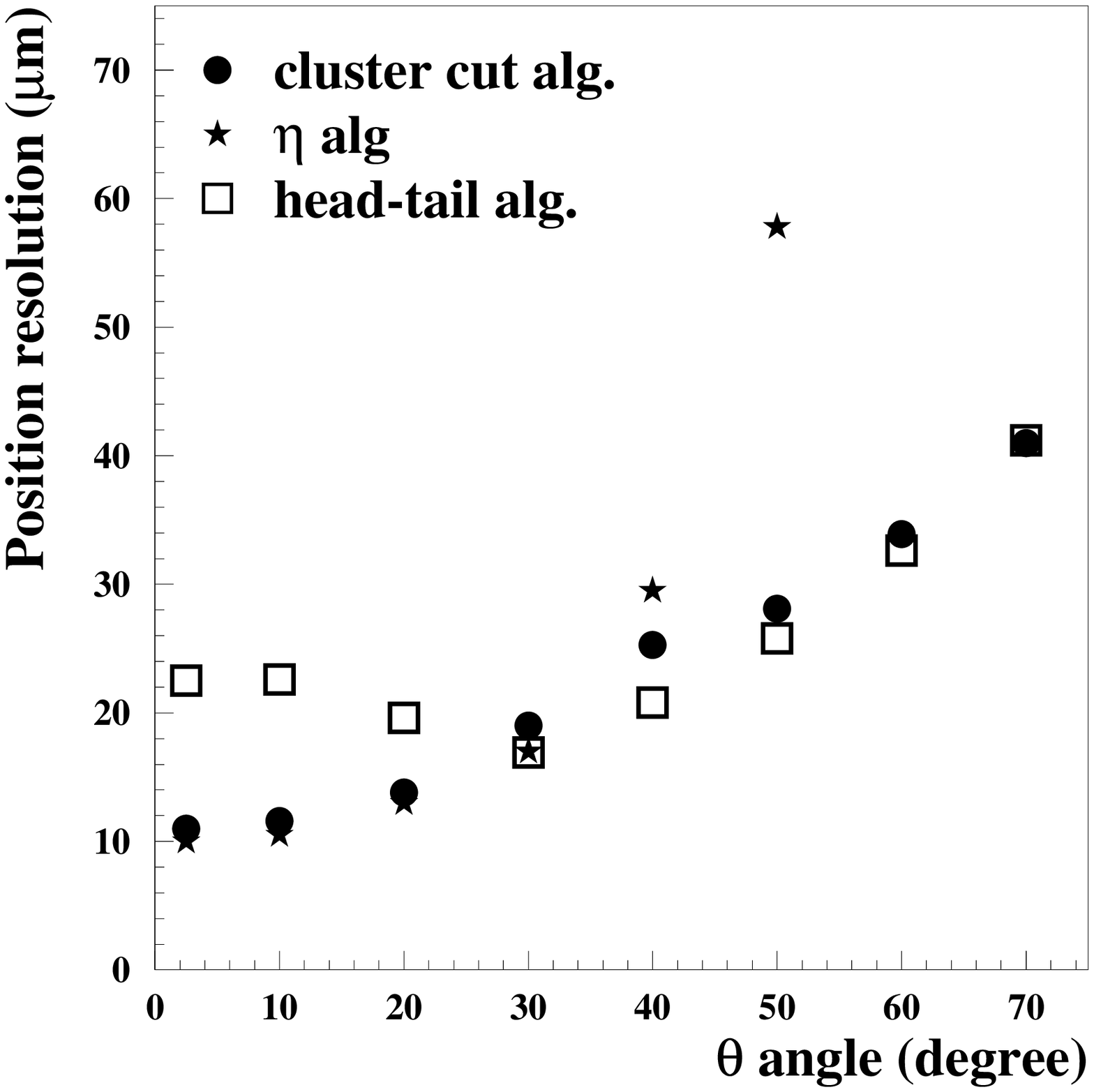,width=8.5cm}
 \put(-15,1.8){(b)}
    }
  \caption[Position resolution as a function of the angle of incidence $\theta$ calculated using 
  general reconstruction algorithms]
  {\rm Position resolution as a function of the angle of incidence $\theta$ calculated using 
    (a) the $\theta_{\rm cut}$ reconstruction algorithm (b) the cluster-cut algorithm. The comparison 
    with the $\eta$ and  head-tail  algorithm is also shown. }
  \label{fig:combang}
\end{figure}

A first simple attempt for a more general algorithm can be based only on the rough knowledge 
of the angle $\theta$ of the incident particle at which the switch between 
two different algorithms is required  ($\theta_{\rm cut}$ algorithm): 
\begin{itemize}
\item if $\theta \lsim \theta_{\rm cut}$ (with $\theta_{\rm cut} \simeq$ 30-40$^\circ$) 
  the uncorrected 3-strips algorithm is used
\item if $\theta > \theta_{\rm cut}$ the head-tail algorithm  is used.
\end{itemize}   
Since the non-linearity correction is not applied, a slightly worse position resolution at small angles is obtained.  
Figure \ref{fig:combang}(a) shows the position resolution obtained with this algorithm as 
function of the incidence angle $\theta$ for two different detectors.

\begin{figure}
\centerline{\epsfig{figure=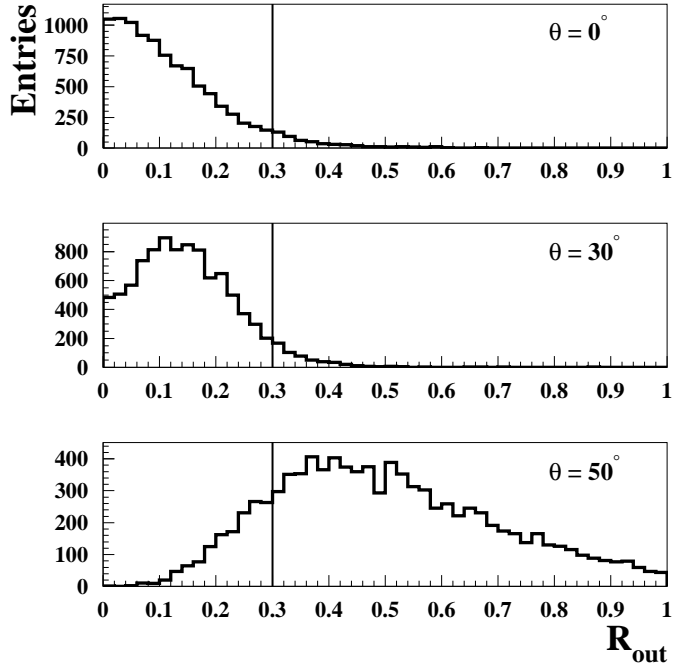,width=10cm}}
  \caption[R$_{\rm out}$ distribution at $\theta = 0^{\circ},\;30^{\circ}, \; 50^{\circ}$]
  {\rm R$_{\rm out}$ distribution at $\theta = 0^{\circ},\;30^{\circ}, \; 50^{\circ}$.  }
  \label{fig:scutplot}
\end{figure}

A more general algorithm based on the use of the cluster size, N$_{\rm strips}$ and a cluster 
shape parameter, R$_{\rm out}$,  is presented in this paper. 
The `cluster cut' algorithm  uses two different calculations as detailed in the following: 
\begin{itemize}
\item if N$_{\rm strips} < $N$_{\rm cut}$
  and R$_{\rm out} < $R$_{\rm cut}$,  where ${\rm R_{out}} = (S_{k-2}+S_{k+2})/S_{k}$,  
  the 3-strips algorithm is used. 
\item if N$_{\rm strips} >$ N$_{\rm cut}$ or 
  R$_{\rm out}  \ge $R$_{\rm cut}$  the head-tail algorithm is employed.
\end{itemize}
To a good approximation the cluster size is directly correlated to the angle of incidence, i.e. small clusters 
(N$_{\rm strips} \sim$ 2-3) are most likely originated by tracks crossing the detector at small angles of incidence whereas 
large clusters (N$_{\rm strips} > 6$) are due to particles crossing the detector at large angles of incidence. 
For intermediate cluster sizes, a matching ambiguity remains. However, the  cluster shape also depends on the angle of incidence:
for larger angles, the track path length is relatively large and thus also the outer strips of smaller clusters have a sizable 
signal (compared to the highest signal in the cluster). 
By making use of this additional information, the ambiguity can be reduced. 
The values  R$_{\rm cut}$=0.3  and N$_{\rm strips}$=5 (same results are obtained for N$_{\rm strips}$=6)  
have been used in this analysis. The cut value for  R$_{\rm cut}$  has been chosen by comparing the distribution for 
different incidence angles (see figure \ref{fig:scutplot}).

Figure \ref{fig:combang}(b) shows the comparison between the position resolution obtained with
the 'cluster cut' algorithm and the  position resolution obtained with the  $\eta$ 
and the head-tail algorithms. The position resolution achieved with the 
uncorrected `cluster-cut' algorithm is only slightly  worse than the one calculated with the 
best standard reconstruction method for each angle (i.e. $\eta$ and  `head-tail').  
Since the `cluster-cut' algorithm does not need  any angular information, it  could be a valuable choice for a first 
position reconstruction  in a general track reconstruction procedure. 

\subsection{Impact of $\delta$-rays on the position resolution}
\label{ss:delta}
\begin{figure}
\begin{center}
\epsfig{file=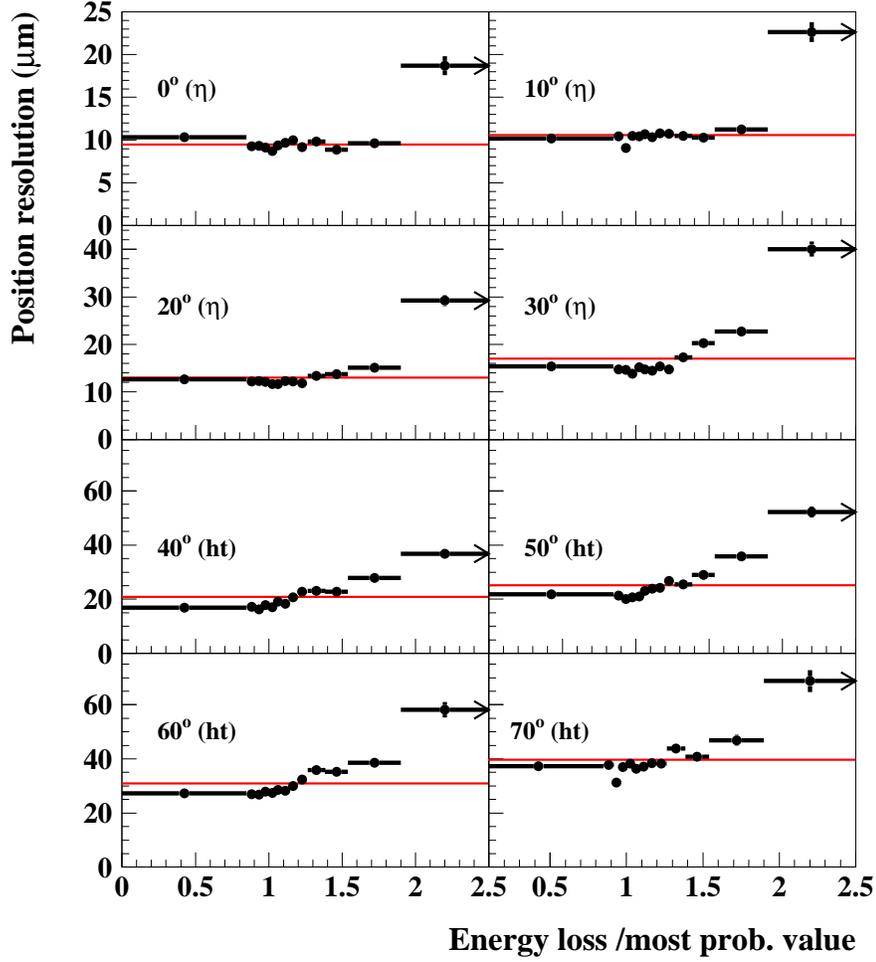,width=13cm}
\end{center}
\caption[Position resolution as a function of the energy loss for several
incident angles]
{\label{deltaangle} Position resolution as a function of the  energy loss
for incident angles 
between $0^\circ$ and  $70^\circ$. The line indicates the average position
resolution.}
\end{figure}

The production of  $\delta$-rays (i.e. knock-on electrons), responsible
for the tail in the energy loss 
distribution, influences the detector resolution,  as it affects the
deposited charge  distribution in 
the silicon bulk. The $\delta$-ray trajectory results in deposition of
charge off the incident 
particle's path  and therefore displaces the charge centre of gravity. 
The position resolution for  larger pulse height becomes significantly
worse as can be seen in  
Figure \ref{deltaangle},  where the width of the residual distribution as a function of the cluster 
pulse height for incident angles between $0^\circ$ and $70^\circ$ is
shown. 
The hit position is reconstructed using the $\eta$-algorithm up to angles
of incidence
of $30^\circ$, whereas for larger incident angles the head-tail algorithm
is used. 
 Although the resolution is worse, events with large signal cluster
cannot be excluded in the tracking reconstruction for the MVD, because
the loss in reconstruction efficiency would be too high. However, it is  
possible to assign different weights to the reconstructed hits when
performing the track fit.

\section {Summary of the results}
\label{sec:summary}

The testbeam setup at DESY has been used to characterise the performance of BMVD
 and FMVD
 detectors plus the prototype readout electronics (HELIX chips). 
The main results from the testbeam measurements 
presented in this paper can be summarised as follows:

\begin{itemize}
\item a signal over noise ratio S/N $\simeq 20$-$24$ has been achieved, the noise level
  is  uniformly distributed over the strips;
\item the detector efficiency $\epsilon$ is very high ($> 99.95\%$);
\item the calibration shows that gain variations of a single HELIX readout chip are of the
  order of 2\% and do not influence the position reconstruction algorithm even in the case of detectors with 
  strips of  different lengths (FMVD);
\item  the charge division has been studied in detail. The
    expected charge sharing between strip implants has been
    confirmed;
\item the charge transfer between
    strip implants and readout strips has been parameterised and
    implemented in a detector simulation program which gives a good description of the data;  
\item the intrinsic position resolution at normal angle of incidence 
  $\sigma_{\rm MVD}^{\rm intr} = 7.2 \pm 0.2 \;\mu {\rm m}$ is highly satisfactory 
  if compared to the value  $p/\sqrt{12} \simeq 35\,\mu$m which represents the limit
    for a digital system with a readout pitch p = 120 $\mu$m;
\item the position resolution is not strongly dependent on the angle $\phi$ in the range of
interest for the FMVD;
\item the position resolution at large angle of incidence $\theta$ is
  still very satisfactory ($\lsim 40~\mu$m up to $\theta = 70 ^\circ$);  
\item a position reconstruction algorithm which uses the rough  knowledge of the
   angle of incidence $\theta$ to choose the most effective reconstruction procedure has been developed. 
  For small incidence angles a 3-strips algorithm is applied whereas for large incidence
  angles the head-tail algorithm is used;
\item a general position reconstruction algorithm, not using prior knowledge 
of the angle of   incidence, has proven to work well up to $\theta = 70 ^\circ$ and could therefore be a 
  valuable choice for a first position reconstruction in the track reconstruction procedure;
\item the production of $\delta$-rays causes a deterioration of
    the position resolution. Since events with very large cluster signal cannot be excluded from the
    MVD data, the hit positions used for a track fit should be
    weighted according to their pulse height.
\end{itemize}



\begin{thebibliography}{99}
%
%
\bibitem{bib:hera_upgrade_scheek}
U.\ Schneekloth (editor), ``The HERA Luminosity Upgrade'', DESY internal
report, DESY-HERA 98-05, 1998.
%
\bibitem{bib:zeus_det}
ZEUS Collaboration, M.\ Derrick et al., The ZEUS Detector, Status
Report 1993, DESY, 1993.
\bibitem{bib:mvd_mech_paper10} A.\ Garfagnini, \Journal{\NIMA}{435}{1999}{34}.
\bibitem{bib:mvd_mech_paper9} R.\ Klanner, ``The ZEUS Micro Vertex Detector'', proceedings of the 
International Europhysics Conference on High Energy Physics, EPS-HEP 99,
Tampere, Finland. 
\bibitem{bib:mvd_mech_paper6} E.\ Koffeman, \Journal{\NIMA}{453}{2000}{89}. 
\bibitem{bib:mvd_mech_paper5} C.\ Coldewey, \Journal{\NIMA}{453}{2000}{149}. 
%
\bibitem{bib:helix_ref1}
 M. Feuerstack-Raible, U. Trunk, et al, ``HELIX 128-x User's Manual Version 2.1, 3.2.1999 '', HD-ASIC-33-0697.  
Available at \\ {\tt \small http://wwwasic.kip.uni-heidelberg.de/\~\,trunk/projects/Helix/}.  
\bibitem{bib:helix_ref2} M.\ Feuerstack-Raible, \Journal{\NIMA}{447}{2000}{35}.
%
\bibitem{bib:mvd_mech_paper3} M.\ C.\ Petruccci, Int. J. Mod. Phys. A Vol. 16 Suppl. 1C (2001) 1078,
\bibitem{bib:mvd_mech_paper2} E.\ Koffeman, \Journal{\NIMA}{473}{2001}{26}.
\bibitem{bib:mvd_mech_paper1} V.\ Chiochia, ``The ZEUS Micro Vertex Detector'', proceedings of the Vertex 2002 conference, hep-ex/0111061,
submitted to \NIMA. 
\bibitem{bib:mvd_mech_paper7} C.\ Coldewey, \Journal{\NIMA}{447}{2000}{44}.
\bibitem{bib:mvd_mech_paper4} A.\ Garfagnini and U.\ K\"otz, \Journal{\NIMA}{461}{2001}{158}. 
\bibitem{bib:mvd_elec_paper}
D. Dannheim et al., {\it Design and Tests of the Silicon Sensors for the ZEUS Micro Vertex Detector}, submitted for
 publication to \NIMA.
%
\bibitem{bib:mvd_mech_paper8} U.\ K\"otz,  \Journal{\NIMA}{461}{2000}{210}. 
\bibitem{bib:labview}
LabView, product of National Instruments.
%
\bibitem{bib:telescope}
C.\ Colledani et al.,
\Journal{\NIMA}{372}{1996}{379}.
%
\bibitem{bib:tele_res}
J.Straver et al.,
\Journal{\NIMA}{348}{1994}{485}.
%
%
%
\bibitem{bib:margherita_phd}
M.\ Milite, PhD Thesis, DESY-THESIS-2001-050.
 Available at \\ 
{\tt http://www-library.desy.de/cgi-bin/showprep.pl?desy-thesis-01-050}.
%
\bibitem{bib:upilex}
ICI Films, High Performance Films Group, Wilmington, Delaware 19897, USA.
%
\bibitem{bib:Gandi}
CERN EST/SM-CI, Photomechanical Technologies Workshop.
%
\bibitem{bib:redondo_phd}
I.\ Redondo, PhD Thesis, DESY-THESIS-2001-037.
 Available at \\
{\tt http://www-library.desy.de/cgi-bin/showprep.pl?desy-thesis-01-037}.
%
\bibitem{bib:padova_helix}
 Internal communication on test of the F/E chips (unpublished).
 Available at {\tt http://zeus.pd.infn.it/MVD/MVD.html}.  
%
\bibitem{bib:moritz_phd}
M.\ Moritz, PhD Thesis, DESY-THESIS-2002-009.
 Available at \\
{\tt http://www-library.desy.de/cgi-bin/showprep.pl?desy-thesis-02-009}
%
\bibitem{bib:simulation} G. Bashindzhagyan and N. Korotkova, {\it Simulation of Silicon Microstrip Detector Resolution for ZEUS Vertex Upgrade}, internal ZEUS note, 99-023, Hamburg (Germany), (1999).
%
%
\bibitem{bib:jan} J. Martens, {\it Simulationen und Qualit$\ddot{a}$tssicherung der Siliziumstreifendetektoren des ZEUS-Mikrovertexdetektors}, DESY-THESIS-1999-044, Diplomarbeit, University of Hamburg
 (1999).\\
%
\bibitem{bib:etaalg}
E.\ Belau et al., 
\Journal{\NIM}{214}{1983}{253}.\\
\bibitem{bib:etaalg1}
R.\ Turchetta, 
\Journal{\NIMA}{335}{1993}{44}.
%
\bibitem{bib:PDG}
 D.E. Groom et al., The European Physical Journal C15 (2000) 1. 
%
\bibitem{bib:headtail}
 KEK Preprint 96-172,
{\it Measurement of the Spatial Resolution of Wide-pitch Silicon Strip Detectors with
 Large Incident Angle}.

%
%
\end{thebibliography}
\end{document}